%%%%%%%%%%%%%%%%%%  tex macros for preprints, cm version %%%%%%%%%%%%%%
%         (P. Ginsparg <ginsparg@lanl.gov>, last updated 7/94)
%         hypertex extensions (still provisional), 7/26/94
%	  Some modifications by C.R.Mafra, 2012

%comment out this line to restore non-hyper functionality
%\input hyperbasics
%
\def\unredoffs{}
\tolerance=1000\hfuzz=2pt
\catcode`\@=11 % This allows us to modify PLAIN macros.
\ifx\hyperdef\UNd@FiNeD\def\hyperdef#1#2#3#4{#4}\def\hyperref#1#2#3#4{#4}\def\href#1#2{#2}\fi
\magnification=1200\unredoffs\baselineskip=16pt plus 2pt minus 1pt
\def\Date#1{\vfill\leftline{#1}\tenpoint\supereject%
\footline={\hss\tenrm\hyperdef\hypernoname{page}\folio\folio\hss}}%
% (restores pagenumbers)

%%%%%% Hour:Minute %%%%%%%%%%%%%%%%%
{\count255=\time\divide\count255 by 60 \xdef\hourmin{\number\count255}
 \multiply\count255 by-60\advance\count255 by\time
 \xdef\hourmin{\hourmin:\ifnum\count255<10 0\fi\the\count255}
}
\def\date{\number\day.\number\month.\number\year\ at \hourmin}

%%%%%%%%%%%% Draft mode %%%%%%%%%%%%%
% puts date/time on each page in big mode, writes labels in margins

% use \nolabels to get rid of eqn, ref, and fig labels in draft mode
\def\nolabels{\def\wrlabeL##1{}\def\eqlabeL##1{}\def\reflabeL##1{}}
\def\writelabels{\def\wrlabeL##1{\leavevmode\vadjust{\rlap{\smash%
{\line{{\escapechar=` \hfill\rlap{\sevenrm\hskip.03in\string##1}}}}}}}%
\def\eqlabeL##1{{\escapechar-1\rlap{\sevenrm\hskip.05in\string##1}}}%
\def\reflabeL##1{\noexpand\llap{\noexpand\sevenrm\string\string\string##1}}}
\nolabels

% tagged sec numbers
\global\newcount\secno \global\secno=0
\global\newcount\meqno \global\meqno=1
\def\s@csym{}

%%%%%%%%% Section %%%%%%%%%%%%%
\def\newsec#1\par{\global\advance\secno by1%
{\toks0{#1}\message{(\the\secno. \the\toks0)}}%
\global\subsecno=0\eqnres@t\let\s@csym\secsym\xdef\secn@m{\the\secno}\noindent
{\bf\hyperdef\hypernoname{section}{\the\secno}{\the\secno.} #1}%
\writetoca{{\string\hyperref{}{section}{\the\secno}{\bf \the\secno\quad}} {\bf #1}}\par%
\nobreak\medskip\nobreak\noindent\ignorespaces}
\def\eqnres@t{\xdef\secsym{\the\secno.}\global\meqno=1\bigbreak\bigskip}
\def\sequentialequations{\def\eqnres@t{\bigbreak}}\xdef\secsym{}

%%%%%%%% Subsection %%%%%%%%%%%
\global\newcount\subsecno \global\subsecno=0
\def\subsec#1\par{\global\advance\subsecno by1%
{\toks0{#1}\message{(\s@csym\the\subsecno. \the\toks0)}}%
\global\subsubsecno=0%
\ifnum\lastpenalty>9000\else\bigbreak\fi
\noindent{\it\hyperdef\hypernoname{subsection}{\secn@m.\the\subsecno}%
{\secn@m.\the\subsecno.} #1}\writetoca{\string\hskip1.45cm
{\string\hyperref{}{subsection}{\secn@m.\the\subsecno}{\secn@m.\the\subsecno.}}
{#1}}\par\nobreak\medskip\nobreak\noindent\ignorespaces}

%%%%%%% Appendix %%%%%%%%%%%%%%
\def\appendix#1#2{\global\meqno=1\global\subsecno=0\xdef\secsym{\hbox{#1.}}%
\bigbreak\bigskip\noindent{\bf Appendix \hyperdef\hypernoname{appendix}{#1}%
{#1.} #2}{\toks0{(#1. #2)}\message{\the\toks0}}%
\xdef\s@csym{#1.}\xdef\secn@m{#1}%
\writetoca{{\string\hyperref{}{appendix}{#1}{\bf {#1}\quad}} {\bf #2}}%
\par\nobreak\medskip\nobreak}

% \eqn\label{a+b=c}   gives displayed equation, numbered consecutively within sections.
% \eqnn, \eqna        define labels in advance, use \eqna\label before an eqalign and
%                     later \label a, \label b etc inside eqalign to get (2.3a), (2.3b) etc
%
\def\checkm@de#1#2{\ifmmode{\def\f@rst##1{##1}\hyperdef\hypernoname{equation}%
{#1}{#2}}\else\hyperref{}{equation}{#1}{#2}\fi}
\def\eqnn#1{\DefWarn#1\xdef #1{(\noexpand\relax\noexpand\checkm@de%
{\s@csym\the\meqno}{\secsym\the\meqno})}%
\wrlabeL#1\writedef{#1\leftbracket#1}\global\advance\meqno by1}
\def\f@rst#1{\c@t#1a\em@ark}\def\c@t#1#2\em@ark{#1}
\def\eqna#1{\DefWarn#1\wrlabeL{#1$\{\}$}%
\xdef #1##1{(\noexpand\relax\noexpand\checkm@de%
{\s@csym\the\meqno\noexpand\f@rst{##1}1}{\hbox{$\secsym\the\meqno##1$}})}
\writedef{#1\numbersign1\leftbracket#1{\numbersign1}}\global\advance\meqno by1}
\def\eqn#1#2{\DefWarn#1%
\xdef #1{(\noexpand\hyperref{}{equation}{\s@csym\the\meqno}%
{\secsym\the\meqno})}$$#2\eqno(\hyperdef\hypernoname{equation}%
{\s@csym\the\meqno}{\secsym\the\meqno})\eqlabeL#1$$%
\writedef{#1\leftbracket#1}\global\advance\meqno by1}
\def\xeqn{\expandafter\xe@n}\def\xe@n(#1){#1}
\def\xeqna#1{\expandafter\xe@n#1}
\def\eqns#1{(\e@ns #1{\hbox{}})}
\def\e@ns#1{\ifx\UNd@FiNeD#1\message{eqnlabel \string#1 is undefined.}%
\xdef#1{(?.?)}\fi{\let\hyperref=\relax\xdef\next{#1}}%
\ifx\next\em@rk\def\next{}\else%
\ifx\next#1\xeqn#1\else\def\n@xt{#1}\ifx\n@xt\next#1\else\xeqna#1\fi
\fi\let\next=\e@ns\fi\next}

\def\DefWarn#1{\ifx\UNd@FiNeD#1\else
\immediate\write16{*** WARNING: the label \string#1 is already defined ***}\fi}
%
% footnotes
\newskip\footskip\footskip14pt plus 1pt minus 1pt %sets footnote baselineskip
\def\footnotefont{\ninepoint}\def\f@t#1{\footnotefont #1\@foot}
\def\f@@t{\baselineskip\footskip\bgroup\footnotefont\aftergroup\@foot\let\next}
\setbox\strutbox=\hbox{\vrule height9.5pt depth4.5pt width0pt}
\global\newcount\ftno \global\ftno=0
\def\foot{\global\advance\ftno by1\def\foot@rg{\hyperref{}{footnote}%
{\the\ftno}{\the\ftno}\xdef\foot@rg{\noexpand\hyperdef\noexpand\hypernoname%
{footnote}{\the\ftno}{\the\ftno}}}\footnote{$^{\foot@rg}$}}
%
%
%     \ref\label{text}
% generates a number, assigns it to \label, generates an entry.
% To list the refs on a separate page,  \listrefs
%
\global\newcount\refno \global\refno=1
\newwrite\rfile
\def\ref{[\hyperref{}{reference}{\the\refno}{\the\refno}]\nref}
\def\nref#1{\DefWarn#1%
\xdef#1{[\noexpand\hyperref{}{reference}{\the\refno}{\the\refno}]}%
\writedef{#1\leftbracket#1}%
\ifnum\refno=1\immediate\openout\rfile=\jobname.refs\fi
\chardef\wfile=\rfile\immediate\write\rfile{\noexpand\item{[\noexpand\hyperdef%
\noexpand\hypernoname{reference}{\the\refno}{\the\refno}]\ }%
\reflabeL{#1\hskip.31in}\pctsign}\global\advance\refno by1\findarg}
%	horrible hack to sidestep tex \write limitation
\def\findarg#1#{\begingroup\obeylines\newlinechar=`\^^M\pass@rg}
{\obeylines\gdef\pass@rg#1{\writ@line\relax #1^^M\hbox{}^^M}%
\gdef\writ@line#1^^M{\expandafter\toks0\expandafter{\striprel@x #1}%
\edef\next{\the\toks0}\ifx\next\em@rk\let\next=\endgroup\else\ifx\next\empty%
\else\immediate\write\wfile{\the\toks0}\fi\let\next=\writ@line\fi\next\relax}}
\def\striprel@x#1{} \def\em@rk{\hbox{}}
\def\lref{\begingroup\obeylines\lr@f}
\def\lr@f#1#2{\DefWarn#1\gdef#1{\let#1=\UNd@FiNeD\ref#1{#2}}\endgroup\unskip}

\def\addref#1{\immediate\write\rfile{\noexpand\item{}#1}} %now unnecessary
\def\listrefs{\vfill\supereject\immediate\closeout\rfile\writestoppt
\baselineskip=\footskip\centerline{{\bf References}}\bigskip{\parindent=20pt%
\frenchspacing\escapechar=` \input \jobname.refs\vfill\eject}\nonfrenchspacing}
\def\startrefs#1{\immediate\openout\rfile=\jobname.refs\refno=#1}
\def\xref{\expandafter\xr@f}\def\xr@f[#1]{#1}
\def\refs#1{\count255=1[\r@fs #1{\hbox{}}]}
\def\r@fs#1{\ifx\UNd@FiNeD#1\message{reflabel \string#1 is undefined.}%
\nref#1{need to supply reference \string#1.}\fi%
\vphantom{\hphantom{#1}}{\let\hyperref=\relax\xdef\next{#1}}%
\ifx\next\em@rk\def\next{}%
\else\ifx\next#1\ifodd\count255\relax\xref#1\count255=0\fi%
\else#1\count255=1\fi\let\next=\r@fs\fi\next}
%

%
% this is ugly, but moore insists
\newwrite\ffile\global\newcount\figno \global\figno=1
\def\fig{fig.~\hyperref{}{figure}{\the\figno}{\the\figno}\nfig}
\def\nfig#1{\DefWarn#1%
\xdef#1{fig.~\noexpand\hyperref{}{figure}{\the\figno}{\the\figno}}%
\writedef{#1\leftbracket fig.\noexpand~\xfig#1}%
\ifnum\figno=1\immediate\openout\ffile=\jobname.figs\fi\chardef\wfile=\ffile%
{\let\hyperref=\relax
\immediate\write\ffile{\noexpand\medskip\noexpand\item{Fig.\ %
\noexpand\hyperdef\noexpand\hypernoname{figure}{\the\figno}{\the\figno}. }
\reflabeL{#1\hskip.55in}\pctsign}}\global\advance\figno by1\findarg}
\def\xfig{\expandafter\xf@g}\def\xf@g fig.\penalty\@M\ {}
\def\figs#1{figs.~\f@gs #1{\hbox{}}}
\def\f@gs#1{{\let\hyperref=\relax\xdef\next{#1}}\ifx\next\em@rk\def\next{}\else
\ifx\next#1\xfig #1\else#1\fi\let\next=\f@gs\fi\next}
%
%% because TeXlive 2011 is buggy wrt to tikz pictures with plain TeX..
\def\figin{\epsfcheck\figin}\def\figins{\epsfcheck\figins}
\def\epsfcheck{\ifx\epsfbox\UnDeFiNeD
\message{(NO epsf.tex, FIGURES WILL BE IGNORED)}
\gdef\figin##1{\vskip2in}\gdef\figins##1{\hskip.5in}% blank space instead
\else\message{(FIGURES WILL BE INCLUDED)}%
\gdef\figin##1{##1}\gdef\figins##1{##1}\fi}
\def\DefWarn#1{}
\def\figinsert{\goodbreak\topinsert}
\def\ifig#1#2#3{\DefWarn#1\xdef#1{fig.~\the\figno}
\writedef{#1\leftbracket fig.\noexpand~\the\figno}%
\figinsert\figin{\centerline{#3}}
\smallskip
\leftskip=20pt \rightskip=20pt
\baselineskip12pt\noindent
{{\bf Fig.~\the\figno}\ \ninepoint #2}
\medskip
\global\advance\figno by1\par\endinsert}
%%%%%%%%%%%%%%%%%%%%%%%%%%%%%%%%%%%%%%%%%%%%%%%%%%%%%%%%%
\newwrite\lfile
{\escapechar-1\xdef\pctsign{\string\%}\xdef\leftbracket{\string\{}
\xdef\rightbracket{\string\}}\xdef\numbersign{\string\#}}
\def\writedefs{\immediate\openout\lfile=label.defs \def\writedef##1{%
{\let\hyperref=\relax\let\hyperdef=\relax\let\hypernoname=\relax
 \immediate\write\lfile{\string\def\string##1\rightbracket}}}}%
\def\writestop{\def\writestoppt{\immediate\write\lfile{\string\pageno
 \the\pageno\string\startrefs\leftbracket\the\refno\rightbracket
 \string\def\string\secsym\leftbracket\secsym\rightbracket
 \string\secno\the\secno\string\meqno\the\meqno}\immediate\closeout\lfile}}
\def\writestoppt{}\def\writedef#1{}

% Section, subsection and appendix labels %
% Note that there must be a blanck line after \newsec,\subsec and before \seclab,\subseclab!
\def\seclab#1{\DefWarn#1%
\xdef #1{\noexpand\hyperref{}{section}{\the\secno}{\the\secno}}%
\writedef{#1\leftbracket#1}\wrlabeL{#1=#1}}
\def\subseclab#1{\DefWarn#1%
\xdef #1{\noexpand\hyperref{}{subsection}{\the\secno.\the\subsecno}%
{\the\secno.\the\subsecno}}\writedef{#1\leftbracket#1}\wrlabeL{#1=#1}}
\def\applab#1{\DefWarn#1%
\xdef #1{\noexpand\hyperref{}{appendix}{\secn@m}{\secn@m}}%
\writedef{#1\leftbracket#1}\wrlabeL{#1=#1}}
\newwrite\tfile \def\writetoca#1{}
\def\leaderfill{\leaders\hbox to 1em{\hss.\hss}\hfill}
% use this to write file with table of contents
\def\writetoc{\immediate\openout\tfile=\jobname.toc
   \def\writetoca##1{{\edef\next{\write\tfile{\noindent ##1
   \string\leaderfill{
% comment this line if you don't want hyperlinked page numbers on TOC
   \string\hyperref{}{page}{\noexpand\number\pageno}%
   {\noexpand\number\pageno}} \par}}\next}}
}
% and this lists table of contents on second pass
\newread\ch@ckfile
\def\listtoc{\immediate\closeout\tfile\immediate\openin\ch@ckfile=\jobname.toc
\ifeof\ch@ckfile\message{no file \jobname.toc, no table of contents this pass}%
\else\closein\ch@ckfile\centerline{\bf Contents}\nobreak\medskip%
{\baselineskip=16pt\footnotefont\parskip=0pt\catcode`\@=11\input\jobname.toc
\catcode`\@=12\bigbreak\bigskip}\fi}
\catcode`\@=12 % at signs are no longer letters
\def\tenpoint{\def\rm{\fam0\tenrm}% switch back to 10-point type
\textfont0=\tenrm \scriptfont0=\sevenrm \scriptscriptfont0=\fiverm
\textfont1=\teni  \scriptfont1=\seveni  \scriptscriptfont1=\fivei
\textfont2=\tensy \scriptfont2=\sevensy \scriptscriptfont2=\fivesy
\textfont\itfam=\tenit \def\it{\fam\itfam\tenit}\def\footnotefont{\ninepoint}%
\textfont\bffam=\tenbf \def\bf{\fam\bffam\tenbf}\def\sl{\fam\slfam\tensl}\rm}
\font\ninerm=cmr9 \font\sixrm=cmr6 \font\ninei=cmmi9 \font\sixi=cmmi6
\font\ninesy=cmsy9 \font\sixsy=cmsy6 \font\ninebf=cmbx9
\font\nineit=cmti9 \font\ninesl=cmsl9 \skewchar\ninei='177
\skewchar\sixi='177 \skewchar\ninesy='60 \skewchar\sixsy='60
\def\ninepoint{\def\rm{\fam0\ninerm}% switch to footnote font
\textfont0=\ninerm \scriptfont0=\sixrm \scriptscriptfont0=\fiverm
\textfont1=\ninei \scriptfont1=\sixi \scriptscriptfont1=\fivei
\textfont2=\ninesy \scriptfont2=\sixsy \scriptscriptfont2=\fivesy
\textfont\itfam=\ninei \def\it{\fam\itfam\nineit}\def\sl{\fam\slfam\ninesl}%
\textfont\bffam=\ninebf \def\bf{\fam\bffam\ninebf}\rm}
%
%---------------------------------------------------------------------
\hyphenation{anom-aly anom-alies coun-ter-term coun-ter-terms}

%%%%%%%%%%%%%%% Subsubsection %%%%%%%%%%%%%%%%%%%%%%%%%%%%%%%%%%%%
\global\newcount\subsubsecno \global\subsubsecno=0
\def\subsubsec#1\par{\global\advance\subsubsecno by1%
{\toks0{#1}\message{(\the\secno\the\subsecno\the\subsubsecno. \the\toks0)}}%
\ifnum\lastpenalty>9000\else\bigbreak\fi
\noindent{\it\hyperdef\hypernoname{subsubsection}{\the\secno.\the\subsecno\the\subsubsecno}%
{\the\secno.\the\subsecno.\the\subsubsecno.} #1}
%%% Add Subsubsections to Index
%\writetoca{\string\quad{\string\hyperref{}{subsubsection}{\the\secno\the\subsecno\the
%\subsubsecno}{\baselineskip=9pt\it\the\secno.\the\subsecno.\the\subsubsecno.}}
% {\baselineskip=9pt\it\ #1}}
\par\nobreak\medskip\nobreak\noindent\ignorespaces}

% Caption for inline tikzpictures
\def\DefWarn#1{}
\def\tikzcaption#1#2{\DefWarn#1\xdef#1{Fig.~\the\figno}
\writedef{#1\leftbracket Fig.\noexpand~\the\figno}%
{
\smallskip
\leftskip=20pt \rightskip=20pt \baselineskip12pt\noindent
{{\bf Fig.~\the\figno}\ \ninepoint #2}
\bigskip
\global\advance\figno by1 \par}}

% convert numbers [1-9] to upper case letters [A-I]
\def\ntoalpha#1{%
\ifcase#1%
@%
\or A\or B\or C\or D\or E\or F\or G\or H\or I
\fi
}

% Appendix label
\global\newcount\appno \global\appno=1
\def\applab#1{\xdef #1{\ntoalpha\appno}\writedef{#1\leftbracket#1}\wrlabeL{#1=#1}
\global\advance\appno by1}

% Clean up the title page definitions
\def\preprint#1 #2\par{\rightline{\vbox{\baselineskip12pt\hbox{#1}\hbox{#2}}}\vskip2cm}
% title with more than one line (note the blanck line in between)
%\title some line
%
%\tile another line
\def\title#1\par{\centerline{\bf #1}\nopagenumbers\pageno=0}
\def\author#1\par{\bigskip\bigskip\centerline{#1}}

\newcount\addressno

\def\email#1#2{\unskip$^#1$\footnote{\null}{\kern-\parindent \llap{$^#1$\hskip1pt}email: #2}}

% centermode for address lines
\def\startcenter{%
  \par
  \begingroup
  \leftskip=0pt plus 1fil
  \rightskip=\leftskip
  \parindent=0pt
  \parfillskip=0pt
}
\def\stopcenter{\endgroup}

\def\address{\bigskip%
  \ifnum\the\addressno=0\else\stopcenter\endgroup\fi
  \advance\addressno by 1%
  \begingroup
  \startcenter
  \it
  \obeylines
  \addressAux
}
\def\addressAux#1{#1}

% need to stop center mode and obeylines from address
\def\abstract{\stopcenter\endgroup\bigskip\bigskip\noindent}

% some sample definitions
\def\Dsl{\,\raise.15ex\hbox{/}\mkern-13.5mu D} %this one can be subscripted
\def\dsl{\raise.15ex\hbox{/}\kern-.57em\partial}
 
\def\boxeqn#1{\vcenter{\vbox{\hrule\hbox{\vrule\kern3pt\vbox{\kern3pt
	\hbox{${\displaystyle #1}$}\kern3pt}\kern3pt\vrule}\hrule}}}

\def\lform{\hbox{$\sqcup$}\llap{\hbox{$\sqcap$}}}
 %pound sterling

\def\l{\lambda}

\def\half{{1\over 2}}

\def\bar{\overline}
\def\({\left(}
\def\){\right)}

 %redefine plain TeX \Im..
%shuffle product

%\owedge

% From Knuth's \pfbox macro
\def\qed{\hbox{\hskip 3pt
%\lower2pt
\vbox{\hrule\hbox to 7pt{\vrule height 7pt\hfill\vrule}
\hrule}}\hskip3pt}

% do not display overfull marks
\overfullrule=0pt\relax

\frenchspacing

% define labels in advance
\newread\instream \openin\instream= label.defs
\ifeof\instream \message{No labels in advance yet. Wait till next pass.}
\else \closein\instream \input label.defs
\fi
\writedefs

%%% References with hyperlinks to arxiv.org; both styles accepted
% Change arXiv to \arXiv ie
% [arXiv:hep-th/1234567].     --> [\arXiv:hep-th/1234567].
% [arXiv:1234.5678 [hep-th]]. --> [\arXiv:1234.5678 [hep-th]].
% Need to strip trailing [hep-th] (if present) to define valid URL
\def\arXiv:#1].{\hepthStrip#1 \nil}
\def\hepthStrip#1 #2\nil{\href{http://arxiv.org/abs/#1}{arXiv:#1 #2\unskip}].}

\input epsf

\def\centretable#1{ \hbox to \hsize {\hfill\vbox{
                    \offinterlineskip \tabskip=0pt \halign{#1} }\hfill} }

\preprint UUITP-38/21

\title Green-Schwarz and Pure Spinor Formulations of Chiral Strings

\author Max Guillen\email{\dagger}{max.guillen@physics.uu.se}

\address
%$^\dagger$ 
Department of Physics and Astronomy, 75108 Uppsala, Sweden

\abstract
Bosonic and RNS chiral strings have been defined from a singular gauge fixing of the respective Polyakov and spinning string actions, enforcing, among other things, the finite nature of their physical spectra. Except for the heterotic case, the tensionless limits of such chiral models have been shown to describe the same field theories predicted by their ambitwistor analogues. In this paper, we study the Green-Schwarz formulation for Type II and heterotic superstrings in a singular gauge. After performing a light-cone gauge analysis, their physical spectra are shown to match those of RNS chiral strings, and their respective tensionless limits are found to describe the same field theories predicted by RNS ambitwistor strings. Their pure spinor counterparts are then introduced by making use of the Oda-Tonin method. In doing so, symmetries hidden in the pure spinor ambitwistor string action become manifest, proposals motivating the sectorized pure spinor BRST charges find simple grounds, and integrated vertex operators emerge naturally.

\Date {August 2021}

\newif\iffig
\figfalse
% to speed up compilation comment the following line
%\input tikz \figtrue

%\def\Box#1,#2,#3,#4,{{\cal N}^{(4)}_{#1|#2,#3,#4}\, I^{(4)}_{#1,#2,#3,#4}}
%\def\Pentagon#1,#2,#3,#4,#5,{{\cal N}^{(5)}_{#1|#2,#3,#4,#5}(\ell) I^{(5)}_{#1,#2,#3,#4,#5}}

%**************************************

\lref\Wittentwistor{
E.~Witten,
``Perturbative gauge theory as a string theory in twistor space,''
Commun. Math. Phys. {\bf 252}, 189-258 (2004).
[arXiv:hep-th/0312171 [hep-th]].
%1179 citations counted in INSPIRE as of 20 Jul 2021
}

\lref\MasonSVA{
  L.~Mason and D.~Skinner,
  ``Ambitwistor strings and the scattering equations,''
JHEP {\bf 1407}, 048 (2014).
[arXiv:1311.2564 [hep-th]].
%%CITATION = arXiv:1311.2564%%
}

\lref\CHYgluonsgravitons{
F.~Cachazo, S.~He and E.~Y.~Yuan,
``Scattering of Massless Particles in Arbitrary Dimensions,''
Phys. Rev. Lett. {\bf 113}, no.17, 171601 (2014).
[arXiv:1307.2199 [hep-th]].
%480 citations counted in INSPIRE as of 20 Jul 2021
}

\lref\CHYeym{
F.~Cachazo, S.~He and E.~Y.~Yuan,
``Einstein-Yang-Mills Scattering Amplitudes From Scattering Equations,''
JHEP {\bf 01}, 121 (2015).
[arXiv:1409.8256 [hep-th]].
%164 citations counted in INSPIRE as of 20 Jul 2021
}

\lref\CHYborninfeld{
F.~Cachazo, S.~He and E.~Y.~Yuan,
``Scattering Equations and Matrices: From Einstein To Yang-Mills, DBI and NLSM,''
JHEP {\bf 07}, 149 (2015).
[arXiv:1412.3479 [hep-th]].
%315 citations counted in INSPIRE as of 20 Jul 2021
}

\lref\Newambis{
E.~Casali, Y.~Geyer, L.~Mason, R.~Monteiro and K.~A.~Roehrig,
``New Ambitwistor String Theories,''
JHEP {\bf 11}, 038 (2015).
[arXiv:1506.08771 [hep-th]].
%104 citations counted in INSPIRE as of 20 Jul 2021
}

\lref\Siegelchiral{
W.~Siegel,
``Amplitudes for left-handed strings,''
[arXiv:1512.02569 [hep-th]].
%44 citations counted in INSPIRE as of 20 Jul 2021
}

\lref\SiegelFactorization{
Y.~t.~Huang, W.~Siegel and E.~Y.~Yuan,
``Factorization of Chiral String Amplitudes,''
JHEP {\bf 09}, 101 (2016).
[arXiv:1603.02588 [hep-th]].
%38 citations counted in INSPIRE as of 20 Jul 2021
}

\lref\Siegelpurespinor{
M.~M.~Leite and W.~Siegel,
``Chiral Closed strings: Four massless states scattering amplitude,''
JHEP {\bf 01}, 057 (2017).
[arXiv:1610.02052 [hep-th]].
%16 citations counted in INSPIRE as of 20 Jul 2021
}

\lref\SiegelCHY{
Y.~Li and W.~Siegel,
``Chiral Superstring and CHY Amplitude,''
[arXiv:1702.07332 [hep-th]].
%10 citations counted in INSPIRE as of 20 Jul 2021
}

\lref\Renannbosonicchiral{
T.~Azevedo, R.~L.~Jusinskas and M.~Lize,
``Bosonic sectorized strings and the $(DF)^{2}$ theory,''
JHEP {\bf 01}, 082 (2020).
[arXiv:1908.11371 [hep-th]].
}

\lref\Renannsectorized{
T.~Azevedo and R.~L.~Jusinskas,
``Connecting the ambitwistor and the sectorized heterotic strings,''
JHEP {\bf 10}, 216 (2017).
[arXiv:1707.08840 [hep-th]].
}

\lref\Chandiatypeii{
O.~Chandia and B.~C.~Vallilo,
``Ambitwistor pure spinor string in a type II supergravity background,''
JHEP {\bf 06}, 206 (2015).
[arXiv:1505.05122 [hep-th]].
}

\lref\CHandiaonshell{
O.~Chandia and B.~C.~Vallilo,
``On-shell type II supergravity from the ambitwistor pure spinor string,''
Class. Quant. Grav. {\bf 33}, no.18, 185003 (2016).
[arXiv:1511.03329 [hep-th]].
%10 citations counted in INSPIRE as of 20 Jul 2021
}

\lref\ChandiaGS {
O.~Chandia and B.~C.~Vallilo,
``Ambitwistor superstring in the Green-Schwarz formulation,''
Eur. Phys. J. C {\bf 77}, no.7, 473 (2017).
[arXiv:1612.01806 [hep-th]].
%3 citations counted in INSPIRE as of 20 Jul 2021
}

\lref\GomezWZA{
  H.~Gomez and E.~Y.~Yuan,
  ``N-point tree-level scattering amplitude in the new Berkovits` string,''
JHEP {\bf 1404}, 046 (2014).
[arXiv:1312.5485 [hep-th]].
%%CITATION = arXiv:1312.5485%%
}

\lref\RenannChiral{
R.~Lipinski Jusinskas,
``Chiral strings, the sectorized description and their integrated vertex operators,''
JHEP {\bf 12}, 143 (2019).
[arXiv:1909.04069 [hep-th]].
%3 citations counted in INSPIRE as of 20 Jul 2021
}

\lref\GreenSchwarz{
M.~B.~Green and J.~H.~Schwarz,
``Covariant Description of Superstrings,''
Phys. Lett. B {\bf 136}, 367-370 (1984).
%770 citations counted in INSPIRE as of 20 Jul 2021
}

\lref\superpoincarequantization{
N.~Berkovits,
``Super Poincare covariant quantization of the superstring,''
JHEP {\bf 04}, 018 (2000).
[arXiv:hep-th/0001035 [hep-th]].
%576 citations counted in INSPIRE as of 20 Jul 2021
}

\lref\GSpurespinors{
N.~Berkovits and D.~Z.~Marchioro,
``Relating the Green-Schwarz and pure spinor formalisms for the superstring,''
JHEP {\bf 01}, 018 (2005).
[arXiv:hep-th/0412198 [hep-th]].
}

\lref\GSpurespinorstwistors{
N.~Berkovits,
``Origin of the Pure Spinor and Green-Schwarz Formalisms,''
JHEP {\bf 07}, 091 (2015).
[arXiv:1503.03080 [hep-th]].
%22 citations counted in INSPIRE as of 20 Jul 2021
}

\lref\purespinorsexplaining{
N.~Berkovits,
``Explaining the Pure Spinor Formalism for the Superstring,''
JHEP {\bf 01}, 065 (2008).
[arXiv:0712.0324 [hep-th]].
%40 citations counted in INSPIRE as of 20 Jul 2021
}

\lref\ICTP{
	N.~Berkovits,
  	``ICTP lectures on covariant quantization of the superstring,''
	[hep-th/0209059].
	%%CITATION = hep-th/0209059%%
}

\lref\InfiniteTension{
  N.~Berkovits,
  ``Infinite Tension Limit of the Pure Spinor Superstring,''
JHEP {\bf 1403}, 017 (2014).
[arXiv:1311.4156 [hep-th], arXiv:1311.4156].
%%CITATION = ICTP-SAIFR-2013-13%%
}

%\cite{Chandia:2021coc}
\lref\chandiabghost{
O.~Chandia and B.~C.~Vallilo,
``Relating the $b$ ghost and the vertex operators of the pure spinor superstring,''
JHEP {\bf 03}, 165 (2021).
[arXiv:2101.01129 [hep-th]].
}

\lref\OdaTonin{
I.~Oda and M.~Tonin,
``On the Berkovits covariant quantization of GS superstring,''
Phys. Lett. B {\bf 520}, 398-404 (2001).
[arXiv:hep-th/0109051 [hep-th]].
%65 citations counted in INSPIRE as of 20 Jul 2021
}

\lref\thetaSYM{
        P.A.~Grassi and L.~Tamassia,
        ``Vertex operators for closed superstrings,''
        JHEP {\bf 0407}, 071 (2004)
        [arXiv:hep-th/0405072].
        %%CITATION = HEP-TH 0405072;%%
}

\lref\NMPS{
	N.~Berkovits,
	``Pure spinor formalism as an N=2 topological string,''
	JHEP {\bf 0510}, 089 (2005).
	[hep-th/0509120].
	%%CITATION = hep-th/0509120%%
}

\lref\Siegelkappa{
W.~Siegel,
``Hidden Local Supersymmetry in the Supersymmetric Particle Action,''
Phys. Lett. B {\bf 128}, 397-399 (1983).
%437 citations counted in INSPIRE as of 21 Jul 2021
}

\lref\supertwistor{
N.~Berkovits, M.~Guillen and L.~Mason,
``Supertwistor description of ambitwistor strings,''
JHEP {\bf 01}, 020 (2020).
[arXiv:1908.06899 [hep-th]].
%6 citations counted in INSPIRE as of 21 Jul 2021
}

\lref\Renannnotes{
R.~L.~Jusinskas,
``Notes on the ambitwistor pure spinor string,''
JHEP {\bf 05}, 116 (2016).
[arXiv:1604.02915 [hep-th]].
%21 citations counted in INSPIRE as of 22 Jul 2021
}

\lref\cederwallborninfeld{
M.~Cederwall and A.~Karlsson,
``Pure spinor superfields and Born-Infeld theory,''
JHEP {\bf 11}, 134 (2011).
[arXiv:1109.0809 [hep-th]].
%19 citations counted in INSPIRE as of 26 Jul 2021
}

\lref\dynamical{
N.~Berkovits,
``Dynamical twisting and the b ghost in the pure spinor formalism,''
JHEP {\bf 06}, 091 (2013).
[arXiv:1305.0693 [hep-th]].
}

\lref\cederwallequations{
N.~Berkovits and M.~Guillen,
``Equations of motion from Cederwall's pure spinor superspace actions,''
JHEP {\bf 08}, 033 (2018).
[arXiv:1804.06979 [hep-th]].
}

\lref\Renannantifield{
R.~L.~Jusinskas,
``Towards the underlying gauge theory of the pure spinor superstring,''
JHEP {\bf 10}, 063 (2019).
[arXiv:1903.10753 [hep-th]].
}

\lref\Thalescurved{
T.~Azevedo and R.~L.~Jusinskas,
``Background constraints in the infinite tension limit of the heterotic string,''
JHEP {\bf 08}, 133 (2016).
[arXiv:1607.06805 [hep-th]].
}

\lref\ChandiaTonin{
O.~Chandia and M.~Tonin,
``BRST anomaly and superspace constraints of the pure spinor heterotic string in a curved background,''
JHEP {\bf 09}, 016 (2007).
[arXiv:0707.0654 [hep-th]].
}

\lref\Toninone{
M.~Tonin,
``Superstrings, K Symmetry and Superspace Constraints,''
Int. J. Mod. Phys. A {\bf 3}, 1519 (1988).
}

\lref\Tonintwo{
M.~Tonin,
``Consistency Condition for Kappa Anomalies and Superspace Constraints in Quantum Heterotic Superstrings,''
Int. J. Mod. Phys. A {\bf 4}, 1983 (1989).
}

\lref\massivepaper{
M.~Guillen, H.~Johansson, R.~L.~Jusinskas and O.~Schlotterer,
``Scattering Massive String Resonances through Field-Theory Methods,''
Phys. Rev. Lett. {\bf 127}, no.5, 051601 (2021).
[arXiv:2104.03314 [hep-th]].
}

\lref\nathanfirstmassive{
N.~Berkovits and O.~Chandia,
``Massive superstring vertex operator in D = 10 superspace,''
JHEP {\bf 08}, 040 (2002).
[arXiv:hep-th/0204121 [hep-th]].
}

\lref\nathanhowe{
N.~Berkovits and P.~S.~Howe,
``Ten-dimensional supergravity constraints from the pure spinor formalism for the superstring,''
Nucl. Phys. B {\bf 635}, 75-105 (2002).
[arXiv:hep-th/0112160 [hep-th]].
}

\lref\Berkovitslize{
N.~Berkovits and M.~Lize,
``Field theory actions for ambitwistor string and superstring,''
JHEP {\bf 09}, 097 (2018).
[arXiv:1807.07661 [hep-th]].
}

\lref\BerkovitsWitten{
N.~Berkovits and E.~Witten,
``Conformal supergravity in twistor-string theory,''
JHEP {\bf 08}, 009 (2004).
[arXiv:hep-th/0406051 [hep-th]].
}

\font\mbb=msbm10 
\newfam\bbb
\textfont\bbb=\mbb

\def\startcenter{%
  \par
  \begingroup
  \leftskip=0pt plus 1fil
  \rightskip=\leftskip
  \parindent=0pt
  \parfillskip=0pt
}
\def\stopcenter{%
  \par
  \endgroup
}

\listtoc
\writetoc
\filbreak

\newsec Introduction

The study of Quantum Field Theories from a worldsheet perspective gained considerable attention after the realization that ${\cal{N}}=4$ super-Yang-Mills can be understood as a string theory living in ${\fam\bbb CP}^{3|4}$ \Wittentwistor. A few years later, Mason and Skinner generalized this twistor construction to arbitrary dimensions by introducing a chiral worldsheet action whose target space coordinates lie in an ambitwistor space \MasonSVA. Although these worldsheet models present vanishing central charge only when the target space dimensions are $D=26$, $D=10$ for the bosonic, supersymmetric cases, respectively, their tree-level correlators were found to remarkably reproduce the CHY formulae for the scattering of gravitons and gluons, valid in arbitrary dimensions \CHYgluonsgravitons. Furthermore, different combinations of matter variables in the ambitwistor string action have been shown to give rise to the several CHY formulae describing tree-level scattering amplitudes of various theories like Born-Infeld, Biadjoint scalar, Galileon, etc \refs{\CHYeym, \CHYborninfeld, \Newambis}. Manifestly supersymmetric formulations of the critical ambitwistor strings were later introduced in \InfiniteTension\ through the use of pure spinor variables.

\medskip
Soon after, Siegel realized that the chiral nature of the ambitwistor string action can actually be understood from a singular gauge-fixing (HSZ) of the Polyakov action \Siegelchiral. Such a singular fixing induces a flip of sign in one of the Virasoro generators of the respective closed string, so that the physical spectrum becomes finite. In this manner, the bosonic chiral string was shown to contain tardyonic, massless and tachyonic states in its physical spectrum, while the RNS chiral string was shown to describe tardyonic and massless states for ${\cal{N}}=1$ supersymmetry, and massless states for ${\cal{N}}=2$ supersymmetry. Using a modified propagator accounting for the singular nature of the gauge choice, computations in conventional string theory were used to calculate several chiral string amplitudes and check they indeed reproduce the CHY/ambitwistor string formulae for massless states \refs{\SiegelFactorization, \Siegelpurespinor, \SiegelCHY}.

\medskip
In recent years, a modification of the pure spinor ambitwistor string has been proposed which closely resembles the left-right splitting in conventional strings \Renannsectorized. This idea was motivated from the observation that the original Type II pure spinor ambitwistor strings coupled to a curved background do not reproduce the Type II supergravity superspace constraints \refs{\nathanhowe,\Chandiatypeii, \CHandiaonshell}. To fix this issue, the authors modified the pure spinor BRST charge by adding an extra term whose flat space limit turned out to be a symmetry, previously ignored, of the pure spinor ambitwistor string. Using dimensional analysis arguments, a dimensionful parameter was reintroduced in the Type II and heterotic pure spinor frameworks, and their physical spectra were shown to coincide with those obtained in RNS chiral strings \Renannsectorized. In this way, it was claimed that these pure spinor modified or sectorized strings are the manifestly supersymmetric versions of the RNS chiral strings. Such an equivalence was later extended to the bosonic case \Renannbosonicchiral\ and, notably, it was shown that the tensionless limit of the bosonic chiral string mixes the different ghost, massless and massive spin-2 states, so that gravitons satisfying $\lform^{3}h = 0$, which were first found in the bosonic ambitwistor string context, naturally fit into the chiral setting. This has given rise to the belief that ambitwistor strings are nothing but the tensionless limit of chiral strings. However, this same phenomenon still has not been proved for the heterotic case. In fact, there exist some subtleties between the pure spinor ambitwistor and sectorized descriptions: The BRST operator of the ambitwistor model proposed in \InfiniteTension\ differs from that constructed in the sectorized model \Renannsectorized. Even though unexpected phenomena might arise in the tensionless limit which make both cohomologies equivalent to each other, this is an open problem to date.

\medskip
On the other hand, it is well known that the Green-Schwarz \GreenSchwarz\ and pure spinor superstrings \superpoincarequantization\ are closely related to each other \refs{\GSpurespinors, \GSpurespinorstwistors, \purespinorsexplaining}. In particular, one can formulate the latter from the former through the Oda-Tonin method \OdaTonin. Morever, since RNS chiral and conventional strings are just different gauge choices of the same action, it is plausible to ask if the standard Green-Schwarz superstring action can account for both types of string theories from different gauge choices. If such a mechanism is indeed feasible, all the accidental features mentioned in the previous paragraph for the pure spinor sectorized string must naturally emerge from it. Although a Green-Schwarz action for Type II ambitwistor strings has been proposed previously in \ChandiaGS, its relation to the familiar string action in conformal gauge is unclear, and the lack of a dimensionful parameter in its construction makes intractable any tensionless limit analysis.

\medskip
In this paper we study the Green-Schwarz superstring action in the singular gauge. The first-order actions obtained from such a gauge-fixing, which we refer to as Green-Schwarz chiral strings, are quantized in light-cone gauge and shown to reproduce the same physical states as those obtained in RNS chiral strings. After adequately taking a tensionless limit, the light-cone gauge equations of motion are shown to coincide with those obtained from RNS ambitwistor strings. A pure spinor worldsheet model is then introduced through the Oda-Tonin method, which reproduces the familiar pure spinor string and ambitwistor string actions after fixing the conformal and singular gauges, respectively. The chiral BRST operators thus constructed match those proposed in the sectorized models. Furthermore, the standard Faddeev-Popov method applied to the pure spinor chiral string introduces b-ghost and delta-function insertions in the path integral which allow us to define integrated vertex operators. Remarkably, the delta-function operator is shown to coincide with the sector-splitting operator recently proposed in \RenannChiral\ from the requirement of BRST-closedness of integrated vertex operators. We finally use the so-called physical operators of 10D super-Yang-Mills \cederwallborninfeld\ to obtain explicit expressions for such vertices.

\medskip
This paper is organized as follows. In section~2, we review the RNS chiral string as a singular gauge choice of the spinning string action, and discuss the Type II and heterotic physical spectra. In section~3, we study the heterotic Green-Schwarz action in the singular gauge, and discuss its light-cone gauge quantization in the tensile and tensionless regimes. A similar analysis is performed for the Type II case in section~4. In section~5, we introduce the pure spinor versions of the Green-Schwarz chiral strings through the Oda-Tonin method, and calculate the integrated vertex operators induced by the respective gauge-fixing procedures. Discussions and further directions are elaborated in section~6. Finally, a light-cone gauge analysis of the RNS heterotic ambitwistor string spectrum is provided in Appendix A.

\newsec Review of RNS Chiral Strings

\seclab\secone

\noindent
In this section we review the concepts of conformal and singular gauges in the Polyakov action, and discuss the RNS chiral strings as well as their physical spectra \Siegelchiral.

\subsec Conformal and Singular Gauges

\subseclab\seconeone

\noindent 
To simplify our analysis, we restrict ourselves to the bosonic case throughout this section. The arguments used here directly apply to the supersymmetric case, which is the subject of study of the next section. The Polyakov action in first-order form reads
\eqnn \bosonicaction
$$ \eqalignno{
S &= \int d^{2}\sigma \bigg[P_{m}\partial_{0}X^{m} + {e \over  {\cal{T}}}(P - {\cal{T}}\partial_{1}X)^{2} + + {\bar{e} \over  {\cal{T}}}(P + {\cal{T}}\partial_{1}X)^{2}\bigg] & \bosonicaction
}
$$
where ${\cal{T}}$ is the string tension, $\sigma^{0}, \sigma^{1}$ are the worldsheet coordinates, $X^{m}$ is the D-dimensional target space coordinate, $P_{m}$ is its respective conjugate momentum, and $e$, $\bar{e}$ are Lagrange multipliers enforcing their respective constraints. The use of the equation of motion for $P_{m}$ allows us to find the relation between the Lagrange multipliers and the worldsheet metric components to be
\eqnn \lagrangemult
$$ \eqalignno{
e &= {1 \over 4 g_{00}}(-\sqrt{-g} + g_{01}) \ , \ \ \ \ \bar{e} = {1 \over 4 g_{00}}(-\sqrt{-g} - g_{01})
& \lagrangemult
}
$$
The usual conformal gauge is then reached by setting $e = \bar{e} = -{{1} \over 4}$. Such a gauge choice actually belongs to a family of metrics which can be parametrized as
\eqnn \metric
$$ \eqalignno{
g_{ij} &= 
\left(\matrix
{
{1\over 2} + \beta & \beta \cr
\beta &  -{1\over 2} + \beta
}
\right) &\metric
}
$$
where $\beta$ is a real number. In this manner, eqn. \lagrangemult\ takes the more compact way
\eqnn \lagrangemulttwo
$$ \eqalignno{
e &= {1 \over 4 }\bigg({-1 -2\beta \over 1 + 2\beta}\bigg) \ , \ \ \ \ \bar{e} =  {1 \over 4 }\bigg({-1  + 2\beta \over 1 + 2\beta}\bigg)
& \lagrangemulttwo
}
$$
One sees from \lagrangemulttwo\ that $\beta = 0$ fixes the conformal gauge. An interesting phenomenon occurs when $\beta \rightarrow \infty$, that is $e = -\bar{e} = -{1 \over 4}$. In such an scenario, the metric determinant vanishes, and so the worldsheet measure gets ill-defined. Moreover, there is no way to integrate $P_{m}$ out in \bosonicaction, and so the action is only defined in a first-order language. For such reasons, this gauge will be referred to as singular gauge. Explicitly, the string action \bosonicaction\ in such a singular gauge takes the form 
\eqnn \singularbosonic
$$ \eqalignno{
S &= \int d^{2}\sigma P^{m}\bar{\partial}X_{m} + \ldots
 & \singularbosonic
}
$$
where $\ldots $ stands for the ghost contributions, and $\bar{\partial} = \partial_{0} + \partial_{1}$. Quantization of \singularbosonic\ tells us that the critical dimension is $D=26$ and the Hilbert space consists of tardyonic and tachyonic spin-2 states whose mass-squares are proportional to ${\cal{T}}$, and the usual massless multiplet made up of the graviton, Kalb-Ramond and dilaton states. Even though this result can be deduced from the explicit computation of the BRST cohomology of \singularbosonic, it can easily be understood from a light-cone gauge perspective: The oscillator algebras for the non-zero modes of the Virasoro constraints possess a relative sign of -1. This implies the normal ordering constants will not longer cancel each other, instead, they will add up to -2. The physical state condition then implies the number operators $N$, $\bar{N}$ must satisfy $N + \bar{N} = 2$. Thus, the three possibilities $(N, \bar{N}) = (0,1,2)$, correspond to the states above mentioned.

\medskip
Remarkably, the null tension limit mixes these three type of spin-2 particles to produce gravitons satisfying the equation of motion $\lform^{3}h = 0$ \Renannbosonicchiral. These are exactly the same gravitons described by the bosonic ambitwistor string \MasonSVA. In this manner, the tensionless limit of the bosonic chiral string reproduces the bosonic ambitwistor string of Mason and Skinner.

\subsec RNS Chiral Strings

\subseclab\seconetwo

The RNS chiral string will be defined by the usual spinning string action in the singular gauge. We discuss the ${\cal{N}}=2$ case in detail and make some comments on the ${\cal{N}}=1$ case at the end of the section.  Explicitly, the ${\cal{N}}=2$ spinning string action is given by
\eqnn \spinningstringaction
$$ \eqalignno{
S &= \int d^{2}\sigma \bigg[P_{m}\partial_{0}X^{m} + \half \psi_{m}\partial_{0}\psi^{m} + \half \bar{\psi}_{m}\partial_{0}\bar{\psi}^{m} + {e \over {\cal{T}}}\bigg[({\cal{P}} - {\cal{T}}\partial_{1}X)^{2} - 2{\cal{T}}\psi_{m}\partial_{1}\psi^{m}\bigg] \cr
& + {\bar{e} \over {\cal{T}}}\bigg[({\cal{P}} + {\cal{T}}\partial_{1}X)^{2} + 2{\cal{T}}\bar{\psi}_{m}\partial_{1}\bar{\psi}^{m}\bigg] + \chi \psi_{m}(P^{m} - \partial_{1}X^{m}) + \bar{\chi} \bar{\psi}_{m}(P^{m} + \partial_{1}X^{m}) \bigg] & \spinningstringaction
}
$$
where $\psi^{m}$, $\bar{\psi}^{m}$ are the ${\cal{N}}=2$ superpartners of $X^{m}$, and $\chi$, $\bar{\chi}$ are Lagrange multipliers enforcing the worldsheet supersymmetry constraints. After setting $e = -\bar{e} = -{1 \over 4}$, $\chi = \bar{\chi} = 0$, the action \spinningstringaction\ becomes
\eqnn \chiralrnsaction
$$\eqalignno{
S &= \int d^{2}\sigma\bigg[P_{m}\bar{\partial}X^{m} + \half \psi_{m}\bar\partial\psi^{m} + \half \bar{\psi}_{m}\bar\partial\bar{\psi}^{m}\bigg] + \ldots
& \chiralrnsaction
}
$$
where $\ldots$ stands for the ghost contributions. Quantization of \chiralrnsaction\ tells us that the critical dimension is $D=10$ and the Hilbert space describes Type II supergravity. The massless nature of the physical spectrum immediately follows from supersymmetry: Indeed, the bosonic and fermionic normal ordering constants cancel each other in each sector, and the physical state condition requires that the number operators satisfy $N+\bar{N} = 0$. The only solution to this equation is when $N=\bar{N} = 0$, and so the space of physical states in this chiral setting is described by the product of two open superstring ground states.

\medskip
Similarly, the ${\cal{N}}=1$ case is quantum-mechanically consistent when $D=10$ for the familiar groups $SO(32)$ and $E_{8}\times E_{8}$. Its physical spectrum is described by ${\cal{N}}=1$ supegravity, ${\cal{N}}=1$ super-Yang-Mills and the first massive level of the open superstring. One more time, this is a direct consequence of supersymmetry: The non-supersymmetric sector acquires a non-zero normal ordering constant. The physical state condition then imposes $N+\bar{N} = 1$ whose solutions $(N,\bar{N}) = (0,1)$ reproduce the physical states just mentioned.

\medskip
These results led us to ask ourselves if similar mechanisms giving rise to the conventional and chiral string theories also take place in the manifestly supersymmetric descriptions of superstrings, that is the Green-Schwarz and pure spinor formulations. We address this problem in the rest of the paper.

\newsec Green-Schwarz Formulation of Heterotic Ordinary and Chiral Strings

\seclab\sectwo

\noindent
In this section we first review the standard description of heterotic Green-Schwarz superstrings in conformal gauge. After imposing the singular gauge, we define the Green-Schwarz version of heterotic chiral strings, and show its light-cone gauge quantization describes the same physical space as that of heterotic RNS chiral strings.

\subsec Heterotic Ordinary Strings

\subseclab\sectwoone

\noindent 
The Green-Schwarz action for heterotic strings is given by \GreenSchwarz
\eqnn\GShetaction
$$\eqalignno{
S &= \int d^{2}\sigma \bigg[{\cal{P}}_{m}\Pi_{0}^{m} + {{\cal{T}}\over 2}\partial_{0}X^{m}(\theta\gamma_{m}\partial_{1}\theta) - {{\cal{T}} \over 2}\partial_{1}X^{m}(\theta\gamma_{m}\partial_{0}\theta) + {e \over {\cal{T}}}({\cal{P}} - {\cal{T}} \Pi_{1})^{2}\cr
 &+ {\bar{e}\over {\cal{T}}}[{(\cal{P}} + {\cal{T}}\Pi_{1})^{2} + T_{J}]\bigg] + S_{J} &\GShetaction
}$$
where $(X^{m}, \theta^{\alpha})$ are the ten-dimensional superspace coordinates, $\Pi_{i}^{m} = \partial_{i}X^{m} + {1\over 2} (\theta\gamma^{m}\partial_{i}\theta)$ for $i=0,1$, $\cal{T}$ stands for the string tension, and $S_{J}$ is the current algebra system. We are using letters from the beginning/middle of the Greek/Latin alphabet to denote $SO(1,9)$ spinor/vector indices. Moreover, $(\gamma^{m})^{\alpha\beta}$, $(\gamma_{m})_{\alpha\beta}$ are the usual $SO(1,9)$ Pauli matrices satisfying $(\gamma^{(m})_{\alpha\beta}(\gamma^{n)})^{\beta\delta} = \eta^{mn}$. The action \GShetaction\ is invariant under the supersymmetry transformations
\eqnn\susyhet
$$\eqalignno{
\delta\theta^{\alpha} &= \epsilon^{\alpha}\ ,  \ \ \ \ \delta X^{m} = -\half(\epsilon\gamma^{m}\theta) \ ,  \ \ \ \ \delta {\cal{P}}^{m} = 0 \ ,  \ \ \ \ \delta e = 0  &\susyhet
}$$
and under the $\kappa$-symmetry transformations \Siegelkappa
\eqnn\kappasymhet
$$\eqalignno{
\delta \theta^{\alpha} = (\gamma^{m}\kappa)^{\alpha}({\cal{P}}_{m} - {\cal{T}}\Pi_{1\,m}) \ ,  \ \ \ \ \delta X^{m} =& \half(\delta\theta\gamma^{m}\theta) \ ,  \ \ \ \
\delta {\cal{P}}^{m} = -{\cal{T}}(\delta\theta\gamma^{m}\partial_{1}\theta)\cr
\delta e = -{\cal{T}}\kappa_{\alpha}\partial_{R}\theta^{\alpha}  \ &,  \ \ \ \ \delta \bar{e} = 0 &\kappasymhet
}$$
where $\partial_{R} = \partial_{0} - 4e\partial_{1}$. The invariance of the action \GShetaction\ under \kappasymhet\ easily follows from the fact that the supersymmetric invariants $\Pi_{i}^{m}$ vary under $\kappa$-symmetry as $\delta \Pi_{i}^{m} = (\delta\theta\gamma^{m}\partial_{i}\theta)$, and thus the full variation of the action reads
\eqnn \variationauxone
$$\eqalignno{
\delta S =& \int d^{2}\sigma \bigg[\delta {\cal{P}}_{m}\Pi_{0}^{m} + {\cal{P}}_{m}(\delta\theta\gamma^{m}\partial_{0}\theta) + {{\cal{T}}\over 4}\bigg(-(\delta\theta\gamma^{m}\theta)(\partial_{0}\theta\gamma_{m}\partial_{1}\theta) + (\delta\theta\gamma^{m}\theta)(\partial_{1}\theta\gamma_{m}\partial_{0}\theta)\bigg)\cr 
& + {\cal{T}}\partial_{0}X^{m}(\delta\theta\gamma_{m}\partial_{1}\theta) - {\cal{T}}\partial_{1}X^{m}(\delta\theta\gamma_{m}\partial_{0}\theta) + {\delta e \over{\cal{T}}}({\cal{P}}-{\cal{T}}\Pi_{1})^{2}  - 4e ({\cal{P}}-{\cal{T}}\Pi_{1})^{m}(\delta\theta\gamma_{m}\partial_{1}\theta)\bigg] \cr
& & \variationauxone
}
$$
After using integration by parts and the 10D identity $(\gamma^{m})_{(\alpha\beta}(\gamma_{m})_{\delta)\epsilon} = 0$, one gets
\eqnn\finalvars
$$\eqalignno{
\delta S &= \int d^{2}\sigma \bigg[({\cal{P}}-{\cal{T}}\Pi_{1})^{m}(\delta\theta\gamma_{m}\partial_{R}\theta) + {\delta e\over {\cal{T}}} ({\cal{P}}-{\cal{T}}\Pi_{1})^{2}\bigg] = 0
& \finalvars
} 
$$
The familiar superstring action is then reached by fixing the conformal gauge: $e = \bar{e} = -{1 \over 4}$, and integrating out ${\cal{P}}^{m}$. Explicitly,
\eqnn\hetactioncg
$$\eqalignno{
S &= {\cal{T}}\int d^{2}\sigma \bigg[{1\over 2}\Pi^{m}\bar{\Pi}_{m} + {1 \over 4}\Pi^{m}(\theta\gamma_{m}\bar{\partial}\theta) - {1 \over 4}\bar{\Pi}^{m}(\theta\gamma_{m}\partial\theta)\bigg] + S_{J} &\hetactioncg
}
$$
where
\eqnn\pisdef
$$\eqalignno{
\Pi^{m} &= \partial X^{m} + \half(\theta\gamma^{m}\partial\theta) \ , \ \ \ \ \bar{\Pi}^{m} = \bar{\partial}X^{m} + \half(\theta\gamma^{m}\bar{\partial}\theta) &\pisdef
}
$$
and $\partial = \partial_{0} - \partial_{1}$, $\bar{\partial} = \partial_{0} + \partial_{1}$. In addition to the Virasoro constraints: $T = \Pi^{m}\Pi_{m}$, $\bar{T} = \bar{\Pi}^{m}\bar{\Pi}_{m} + T_{J}$, the action \hetactioncg\ is subject to the constraints
\eqnn\dcg
$$\eqalignno{
d_{\alpha} &= p_{\alpha} - \half(\gamma^{m}\theta)_{\alpha}\Pi_{m} - {1\over 4}(\gamma^{m}\theta)_{\alpha}(\theta\gamma_{m}\partial_{1}\theta) & \dcg
}
$$
as can easily be seen from the standard definition of the momentum variable $p_{\alpha} = {\delta{\cal{L}} \over \delta \partial_{0}\theta^{\alpha}}$. Using standard Poisson brackets, these constraints can be shown to satisfy the SUSY-like algebra
\eqnn\susylikeal
$$\eqalignno{
\{d_{\alpha}(\sigma), d_{\beta}(\sigma')\} &= -(\gamma^{m})_{\alpha\beta}\Pi_{m}(\sigma')\delta(\sigma-\sigma') &\susylikeal
}
$$
Since $\Pi^{m}$ is null, $d_{\alpha}$ describes 8 first-class and 8 second-class constraints. As is very well known, there is not simple way to separate out both types of constraints in a Lorentz covariant manner. However, one can use the symmetries of \GShetaction\ to fix the so-called light-cone gauge and study the physical spectrum of the model. Such an analysis can be found in any standard string theory textbook, and thus the details will be omitted here.

\subsec Heterotic Chiral Strings

\subseclab\sectwotwo

One might ask if there exists a mechanism similar to that studied in section \secone\ which gives rises to a string model whose physical spectrum matches that of the heterotic RNS chiral string. This motivates the gauge choice $e= -\bar{e}= -{1\over 4}$ in \GShetaction, which yields the action
\eqnn\almostchiralaction
$$\eqalignno{
S &= \int d^{2}\sigma \bigg[{\cal{P}}_{m}\bar{\Pi}^{m} + {{\cal{T}} \over 4}\Pi^{m}(\theta\gamma_{m}\bar{\partial}\theta) - {{\cal{T}}\over 4}\bar{\Pi}^{m}(\theta\gamma_{m}\partial\theta)\bigg] + S_{J} &\almostchiralaction
}
$$
Eqn. \almostchiralaction\ is a first-order action in the bosonic coordinates. Indeed, the conjugate momentum associated to $X^{m}$ can easily be computed to be
\eqnn \newmomentum
$$ \eqalignno{
P_{m} &= {\cal{P}}_{m} + {{\cal{T}} \over 2} (\theta\gamma_{m}\partial_{1}\theta)
& \newmomentum
}
$$
Eqns \susyhet, \kappasymhet, \newmomentum\ then imply that the momentum $P_{m}$ varies under SUSY as
\eqnn \newtransformationsone
$$ \eqalignno{
\delta P_{m} &= {{\cal{T}} \over 2}(\epsilon \gamma_{m}\partial_{1}\theta)  & \newtransformationsone
}
$$
and under $\kappa$-transformations as
\eqnn \newtransformationstwo
$$ \eqalignno{
\delta P_{m} &= -{{\cal{T}} \over 2}\partial_{1}(\delta\theta\gamma_{m}\theta) & \newtransformationstwo
}
$$
and so, unlike the pair $(X^{m}, {\cal{P}}_{m})$, the phase space bosonic coordinates $(X^{m}, P_{m})$ are similarly treated by the symmetries of \almostchiralaction. These subtleties are obscure in the conformal gauge, since the variable ${\cal{P}}^{m}$ can be integrated out. However, the singular gauge is intrinsically related to a first-order formulation, and so a description containing $P_{m}$, instead of ${\cal{P}}_{m}$, turns out to be more convenient for its study. Having said this, we introduce the supersymmetric invariants
\eqnn\susyobjectsone
\eqnn\susyobjectstwo
\eqnn\susyobjectsthree
$$ \eqalignno{
{\cal{P}}_{L}^{m} &= P^{m} - {\cal{T}}\partial_{1} X^{m} - {\cal{T}}(\theta\gamma^{m}\partial_{1}\theta)& \susyobjectsone\cr
{\cal{P}}_{R}^{m} &= P^{m} + {\cal{T}}\partial_{1} X^{m}& \susyobjectstwo\cr
{\cal{P}}^{m} &= \half({\cal{P}}_{L}^{m} + {\cal{P}}_{R}^{m}) & \susyobjectsthree
}
$$
and rewrite the Green-Schwarz action \GShetaction\ in the form
\eqnn\GShetactionalt
$$ \eqalignno{
S &= \int d^{2}\sigma \bigg[{\cal{P}}_{m}\Pi_{0}^{m} + {{\cal{T}}\over 2}\partial_{0}X^{m}(\theta\gamma_{m}\partial_{1}\theta) - {{\cal{T}}\over 2}\partial_{1}X^{m}(\theta\gamma_{m}\partial_{0}\theta) + {e \over {\cal{T}}}{\cal{P}}_{L}^{2}\cr
& + {\bar{e}\over {\cal{T}}}[{\cal{P}}_{R}^{2} + T_{J}]\bigg] + S_{J} & \GShetactionalt
}
$$
The invariance of \GShetactionalt\ under SUSY and $\kappa$-symmetry immediately follows from the invariance of \GShetaction\ under both symmetries. For instance, the $\kappa$-transformations\ \kappasymhet, \newtransformationstwo, imply the supersymmetric invariants vary as
\eqnn\l
$$ \eqalignno{
\delta{\cal{P}}_{L}^{m} = -2{\cal{T}}(\delta\theta\gamma^{m}\partial_{1}\theta) \ &, \ \ \ \ 
 \delta{\cal{P}}_{R}^{m} = 0 \cr
\delta{\cal{P}}_{m} = -{\cal{T}}(\delta\theta\gamma^{m}\partial_{1}\theta) \ &, \ \ \ \ 
\delta \Pi_{i}^{m} = (\delta\theta\gamma^{m}\partial_{i}\theta) & \l
}
$$
and so the variation of \GShetactionalt\ can be written as
\eqnn\varone
$$ \eqalignno{
\delta S &= \int d^{2}\sigma\, \bigg[({\cal{P}}^{m}-{\cal{T}}\Pi_{1}^{m})(\delta\theta\gamma_{m}\partial_{0}\theta) + {\delta e \over {\cal{T}}} {\cal{P}}_{L}^{2} - 4e{\cal{P}}_{L}^{m}(\delta\theta\gamma_{m}\partial_{1}\theta)\bigg] & \varone
}
$$
Using that ${\cal{P}}_{L}^{m} = {\cal{P}}^{m} - {\cal{T}}\Pi_{1}^{m}$, and $\partial_{R} = \partial_{0} - 4e\partial_{1}$, one readily concludes that
\eqnn\vartwo
$$ \eqalignno{
\delta S &= \int d^{2}\sigma\, \bigg[{\cal{P}}_{L}^{m}(\delta\theta\gamma_{m}\partial_{R}\theta) + {\delta e\over {\cal{T}}} {\cal{P}}_{L}^{2}\bigg] = 0 & \vartwo
}
$$
The singular gauge $e= -\bar{e} = - {1 \over 4}$ then yields
\eqnn\GShetchiralalt
$$ \eqalignno{
S &= \int d^{2}\sigma\, \bigg[P_{m}\bar{\Pi}^{m} + {{\cal{T}}\over 4}(\Pi^{m}- \bar{\Pi}^{m})(\theta\gamma_{m}\bar{\partial}\theta)\bigg] +  S_{J}
& \GShetchiralalt\
}
$$
Although this action is not conformally invariant, its light-cone gauge and pure spinor versions will be. As will be seen in the next section, the physical spectrum of \GShetchiralalt\ coincides with that of heterotic RNS chiral strings, and thus \GShetchiralalt\ will be referred to as the heterotic Green-Schwarz chiral string action. In addition to the Virasoro-like constraints $T = {\cal{P}}^{m}_{L}{\cal{P}}_{L\,m}$, $\bar{T} = {\cal{P}}_{R}^{m}{\cal{P}}_{R\,m} + T_{J}$, the system \GShetchiralalt\ is also subject to the fermionic constraints
\eqnn\dalt
$$ \eqalignno{
d_{\alpha} &= p_{\alpha} - \half(\gamma^{m}\theta)_{\alpha}P_{m} + {{\cal{T}}\over 2}(\gamma^{m}\theta)_{\alpha}\Pi_{1\,m}\cr
 &= p_{\alpha} - \half(\gamma_{m}\theta)_{\alpha}{\cal{P}}_{L}^{m} - {{\cal{T}}\over 4}(\gamma_{m}\theta)_{\alpha}(\theta\gamma^{m}\partial_{1}\theta) & \dalt
}
$$
which satisfy the SUSY-like algebra
\eqnn \niceds
$$ \eqalignno{
\{d_{\alpha}(\sigma), d_{\beta}(\sigma')\} &= -(\gamma^{m})_{\alpha\beta}{\cal{P}}_{L\,m}\delta(\sigma-\sigma') & \niceds
}
$$
As in the ordinary superstring, eqn. \niceds\ allows us to easily conclude this system will have 8 first-class and 8 second-class constraints. In particular, this implies standard light-cone gauge quantization techniques can be used to study the physical spectrum. This is what we do next.

\subsec Light-Cone Gauge Quantization

\subseclab\sectwothree

The light-cone gauge is reached by using the $\kappa$-symmetry transformations \kappasymhet\ to impose that $(\gamma^{+}\theta)_{\alpha} = 0$, and the Virasoro-like transformations given by
\eqnn \virasoroliketransf
$$ \eqalignno{
\delta_{\xi_{L}} P^{m} &= -\partial_{1}\xi_{L} {\cal{P}}^{m}_{L} - \xi_{L}\partial_{1}{\cal{P}}^{m}_{L} \ , \ \ \ \ \delta_{\xi_{L}} X^{m} = {\xi_{L} {\cal{P}}_{L}^{m}\over {\cal{T}}} \cr
\delta_{\xi_{R}} P^{m} &= \partial_{1}\xi_{R} {\cal{P}}^{m}_{R} + \xi_{R}\partial_{1}{\cal{P}}^{m}_{R} \ \ \ , \ \ \ \ \delta_{\xi_{R}} X^{m} = {\xi_{R} {\cal{P}}_{R}^{m}\over {\cal{T}}} & \virasoroliketransf
}
$$
to impose that $P^{+} = k^{+}$, $X^{+} = 0$, where $k^{+}$ is a constant. Notice that a similar gauge choice was used in \supertwistor\ to construct a light-cone gauge description of RNS ambitwistor strings. The gauge-fixed action then takes the form
\eqnn \lcgaction
$$ \eqalignno{
S &= \int d^{2}\sigma \bigg[P^{i}\bar{\partial}X^{i} + S^{a}\bar{\partial}S^{a}\bigg]  + S_{J} & \lcgaction
}
$$
where we have introduced the variables $S^{a} = \sqrt{\sqrt{2}k^{+}}\theta^{a}$, and the symbols $a$, $\dot{a}$ denote $SO(8)$ chiral, antichiral spinor indices, respectively. The equations of motion following from \lcgaction\ imply the worldsheet fields are chiral, and thus their mode expansions can be cast as
\eqnn \modesone
$$ \eqalignno{
X^{i} &= \sum_{n=-\infty}^{+\infty}x_{n}^{i}e^{in(\sigma_{0}-\sigma_{1})} \ , \ \ \ \ P^{i} = \sum_{n=-\infty}^{+\infty} p^{i}_{n}e^{in(\sigma_{0}-\sigma_{1})} \ , \ \ \ \ S^{a} = \sqrt{i}\sum_{n=-\infty}^{+\infty}S^{a}_{n}e^{in(\sigma_{0} - \sigma_{1})} \cr 
& & \modesone
} 
$$
where the numerical factor in front of the fermionic mode expansion was chosen for convenience. The standard commutation relations
\eqnn \commutatatorsone
$$ \eqalignno{
[P^{m}(\sigma), X^{n}(\sigma')] &= i\eta^{mn}\delta(\sigma - \sigma') \ , \ \ \ \  \{S^{a}(\sigma), S^{b}(\sigma')\} = i\delta^{ab}\delta(\sigma-\sigma')
& \commutatatorsone
}
$$
then provide the familiar mode algebras
\eqnn \pxs
$$ \eqalignno{
[p_{m}^{i}, x_{n}^{j}] &= i\delta_{m,-n}\eta^{ij} \ , \ \ \ \ \{S^{a}_{m}, S^{b}_{n}\} = \delta_{m,-n}\delta^{ab}
& \pxs
}
$$
As usual, the Hilbert space is spanned by states annihilated by the Virasoro-like constraints $T = {\cal{P}}_{L}^{m}{\cal{P}}_{L\,m}$, $\bar{T} = {\cal{P}}_{R}^{m}{\cal{P}}_{R\,m} + T_{J}$. In particular, their zero modes defined as
\eqnn \zeromodesvir
$$ \eqalignno{
T_{0} &= \int_{0}^{2\pi} d\sigma_{1} T(\sigma_{0}, \sigma_{1}) \ , \ \ \ \ \bar{T}_{0} = \int_{0}^{2\pi} d\sigma_{1} \bar{T}(\sigma_{0},\sigma_{1}) & \zeromodesvir
}
$$
must annihilate any physical state. To interpret the physical meaning of eqn. \zeromodesvir, we first need to write down the Virasoro-like constraints in light-cone gauge. After a few simple algebraic manipulations, one finds that
\eqnn \auxtwo
$$ \eqalignno{
T &= -2k^{+}[P^{-} - {\cal{T}}\partial_{1}X^{-} + {{\cal{T}}\over k^{+}}S^{a}\partial_{1}S^{a}] + (P^{i} - {\cal{T}}\partial_{1}X^{i})(P^{i} - {\cal{T}}\partial_{1}X^{i})\cr
\bar{T} &= -2k^{+}[P^{-} + {\cal{T}}\partial_{1}X^{-}] + (P^{i} + {\cal{T}}\partial_{1}X^{i})(P^{i} + {\cal{T}}\partial_{1}X^{i}) + T_{J} & \auxtwo
}
$$
and so eqn. \zeromodesvir\ takes the form
\eqnn \auxthree
$$ \eqalignno{
T_{0} &= -2p_{0}^{-}k^{+} + 2{\cal{T}}\bigg[-\sum_{n=-\infty}^{+\infty}nS^{a}_{-n}S^{a}_{n} + \sum_{n=-\infty}^{+\infty}\alpha^{i}_{-n}\alpha^{i}_{n}\bigg]\cr
\bar{T}_{0} &= -2p_{0}^{-}k^{+} + 2{\cal{T}}\bigg[\sum_{n=-\infty}^{+\infty}\beta^{i}_{-n}\beta^{i}_{n} + \bar{N}_{J}\bigg] & \auxthree
}
$$
where $\bar{N}_{J}$ is the oscillator sum associated to the current algebra system, and the oscillators $\alpha_{n}^{i}$, $\beta_{n}^{i}$ are defined as
\eqnn \defalphabeta
$$ \eqalignno{
\alpha_{n}^{i} &= {1 \over \sqrt{2{\cal{T}}}}[p_{n}^{i} + in{\cal{T}}x_{n}^{i}] \ , \ \ \ \ \beta_{n}^{i} = {1 \over \sqrt{2{\cal{T}}}}[p_{n}^{i} - in{\cal{T}}x_{n}^{i}] & \defalphabeta
}
$$
and thus satisfy the following algebras
\eqnn \algebraalphabeta
$$ \eqalignno{
[\alpha_{m}^{i}, \alpha_{n}^{j}] &= -n\delta_{m,-n}\delta^{ij} \, \ \ \ \ [\beta_{m}^{i}, \beta_{n}^{j}] = n\delta_{m,-n}\delta^{ij} & \algebraalphabeta
}
$$
The following convenient redefinitions
\eqnn \redefalphabeta
$$ \eqalignno{
\alpha^{i}_{n} &\rightarrow i\sqrt{|n|}\tilde{\alpha}^{i}_{n} \ , \ \ \ \  \beta^{i}_{n} \rightarrow \sqrt{|n|}\tilde{\beta}^{i}_{n} & \redefalphabeta
}
$$
for $n\neq 0$, then allow us to write the zero modes $T_{0}$, $\bar{T}_{0}$ as
\eqnn \Tfinal
\eqnn \barTfinal
$$ \eqalignno{
T_{0} &= -2p_{0}^{-}k^{+} + 2{\cal{T}}\alpha^{i}_{0}\alpha^{i}_{0} - 4{\cal{T}}\bigg[\sum_{n=1}^{+\infty}n:S^{a}_{-n}S^{a}_{n}: + \sum_{n=1}^{+\infty}n:\tilde{\alpha}^{i}_{-n}\tilde{\alpha}^{i}_{n}: + n_{0}\bigg] & \Tfinal \cr
\bar{T}_{0} &= -2p_{0}^{-}k^{+} + 2{\cal{T}}\beta^{i}_{0}\beta^{i}_{0} + 4{\cal{T}}\bigg[\sum_{n=1}^{+\infty}n:\tilde{\beta}^{i}_{-n}\tilde{\beta}^{i}_{n}: + \bar{N}_{J} +\bar{n}_{0}\bigg] & \barTfinal
}
$$
where $:\,:$ stands for normal ordering, $n_{0}$, $\bar{n}_{0}$ are zero-point energies, and the oscillators appearing in eqns. \Tfinal, \barTfinal\ obey the standard harmonic oscillator algebras
\eqnn \harmonicalgebras
$$ \eqalignno{
[\tilde{\alpha}^{i}_{m}, \tilde{\alpha}^{j}_{n}] &= \delta_{m,-n}\delta^{ij} \ , \ \ \ \ [\tilde{\beta}^{i}_{m}, \tilde{\beta}^{j}_{n}] = \delta_{m,-n}\delta^{ij} \ , \ \ \ \ \{S^{a}_{m}, S^{b}_{n}\} = \delta_{m,-n}\delta^{ab} & \harmonicalgebras
}
$$
The constants $n_{0}$, $\bar{n}_{0}$ can be obtained in the very same way as they are in ordinary strings. In this way, $n_{0} = 0$ since bosonic and fermionic contributions cancel each other, and $\bar{n}_{0} = -{8\over 24} -{32\over 48} = -1$ for the gauge groups $SO(32)$ and $E_{8} \times E_{8}$. Therefore, one learns that
\eqnn \Tfinalfinal
\eqnn \barTfinalfinal
$$ \eqalignno{
T_{0} &= -2p_{0}^{-}k^{+} + p^{i}_{0}p^{i}_{0} - 4{\cal{T}}\bigg[N_{S} + N_{\tilde{\alpha}}\bigg] & \Tfinalfinal\cr
\bar{T}_{0} &= -2p_{0}^{-}k^{+} + p^{i}_{0}p^{i}_{0} + 4{\cal{T}}\bigg[\bar{N}_{\tilde{\beta}} + \bar{N}_{J} - 1\bigg] & \barTfinalfinal
}
$$
where $N_{\tilde{\alpha}}$, $\bar{N}_{\tilde{\beta}}$, $N_{S}$, $\bar{N}_{J}$ are the number operators corresponding to the oscillator systems in eqn. \harmonicalgebras\ and the current algebra sector. Considering that physical states are momentum eigenvectors, eqns. \Tfinalfinal, \barTfinalfinal\ impose the following restrictions
\eqnn \massformulaone
\eqnn \massformulatwo
$$ \eqalignno{
k^{2} - 4{\cal{T}}\bigg[N_{S} + N_{\tilde{\alpha}}\bigg] &= 0 & \massformulaone\cr
k^{2} + 4{\cal{T}}\bigg[\bar{N}_{\tilde{\beta}} + \bar{N}_{J} - 1\bigg] &= 0 & \massformulatwo
}
$$
Consistency of eqns. \massformulaone, \massformulatwo\ then requires that

\eqnn \nequations
$$ \eqalignno{
N_{S} + N_{\tilde{\alpha}} + \bar{N}_{\tilde{\beta}} + \bar{N}_{J} &= 1 & \nequations
}
$$
and so the spectrum is finite. 

\medskip
As is well known, the vacuum is a degenerate state which realizes the fermionic zero mode algebra \pxs, namely
\eqnn \auxfour
$$ \eqalignno{
\{S_{0}^{a}, S_{0}^{b}\} &= \delta^{ab} & \auxfour
}
$$
It is described by an $SO(8)$ vector $\psi_{i}$, and an $SO(8)$ antichiral spinor $\psi_{\dot{a}}$ obeying
\eqnn \auxfive
$$ \eqalignno{
S_{0,a}\psi_{i} &= {1\over \sqrt{2}}(\sigma_{i})_{a\dot{a}}\psi^{\dot{a}} \cr
S_{0,a}\psi_{\dot{a}} &= {1\over \sqrt{2}}(\sigma_{i})_{a\dot{a}}\psi^{i} & \auxfive
}
$$
where $(\sigma^{i})_{a\dot{a}}$ are the $SO(8)$ Pauli matrices. For the sake of simplicity, we will use the notation $\psi_{0,k} = \psi^{i} \otimes \psi^{\dot{a}}$, to denote such a degenerate vacuum with momentum $k^{m}$. Therefore, the physical spectrum contains massless states describing ${\cal{N}}=1$ super-Yang-Mills and ${\cal{N}}=1$ supergravity, and a single massive state which coincides with the first massive level of the open superstring, as shown in Table 1. This is the same physical spectrum of the heterotic RNS chiral string discussed in section \seconetwo.

\topinsert

\centerline{\vbox{%\offinterlineskip
\tabskip=0pt
\halign{ 
\vrule height2.75ex depth1.25ex width 0.6pt #\tabskip=1em &
\hfil #\hfil &\vrule # & \qquad # \hfil &\vrule # &
\hfil  #\hfil &#\vrule width 0.6pt \tabskip=0pt\cr
\noalign{\hrule height 0.6pt}
&  Mass Level && \hfil Physical States\hfil  &&  Number of States  &\cr
\noalign{\hrule}
& $M^{2} = 0$ && $(\psi_{J}^{i}\oplus \psi_{J}^{\dot{a}})$, $\tilde{\beta}^{j}_{-1}(\psi^{i}\oplus \psi^{\dot{a}})$ && 72B$\,+\,$72F &\cr
& $M^{2} = 4{\cal{T}}$ && $\tilde{\alpha}^{j}_{-1}(\psi^{i}\oplus \psi^{\dot{a}})$,  $S^{a}_{-1}(\psi^{i}\oplus \psi^{\dot{a}})$ && 128B$\,+\,$128F &\cr
\noalign{\hrule height 0.6pt}
}}}
\smallskip
{\leftskip=20pt\rightskip=20pt\baselineskip6pt\noindent\ninepoint
{\bf Table 1.} ${\cal{N}}=1$ SYM $+$ SUGRA, and the 1st. massive level of the open superstring from the quantization of heterotic chiral strings.
\par}
\endinsert

\subsec Tensionless Limit

\subseclab\sectwofour

When the string tension becomes null, the generators ${\cal{P}}^{2}_{L}$, ${\cal{P}}^{2}_{R}$ are not longer independent of each other. However, the respective Hilbert space is still generated by two independent objects, namely $P_{m}$, $\partial X^{m}$. To account for it, the physical states studied in the previous section need to be expressed in the basis $(P_{m}, \partial X^{m})$ before taking the limit ${\cal{T}} \rightarrow 0$. Let us illustrate this idea with the gravitational states. The heterotic chiral string gravitons are given by
\eqnn \tensionlessgrav
$$ \eqalignno{
G_{ij} &=  \tilde{h}_{ij} + h_{ij} \ , \ \ \ \ \tilde{G}_{ij} = {\cal{T}}(\tilde{h}_{ij} - h_{ij})
& \tensionlessgrav
}
$$
when written in the basis $(P_{m}, \partial X^{m})$, where $\tilde{h}_{ij}$ is the massive graviton satisfying $\lform\tilde{h}_{ij} = {\cal{T}}\tilde{h}_{ij}$, and $h_{ij}$ is the supergravity graviton satisfying $\lform h_{ij} = 0$. It is not hard to see that the fields in \tensionlessgrav\ satisfy the equations of motion
\eqnn \newgravitons
$$ \eqalignno{
2\lform G_{ij} &= {\cal{T}}G_{ij} + \tilde{G}_{ij} \ , \ \ \ \  \lform \tilde{G}_{ij} = {\cal{T}}\lform G_{ij} & \newgravitons
}
$$
so that when ${\cal{T}} \rightarrow 0$, one obtains
\eqnn \newgravitonstwo
$$ \eqalignno{
\tilde{G}_{ij} &= 2\lform G_{ij} \ , \ \ \ \  \lform^{2} G_{ij} = 0& \newgravitonstwo
}
$$
As discussed in Appendix A, these equations have the same form as those found in the heterotic ambitwistor string when written in light-cone gauge. Furthermore, a similar analysis shows us that the 2-form fields of the tensionless chiral and ambitwistor strings obey a similar relation as that in \newgravitonstwo. By supersymmetry, the equivalence extends to the fermionic states. These all results together with the fact the number of degrees of freedom in both theories can be demonstrated to be 384, 192 bosons and 192 fermions, allow us to conclude the tensionless limit of the heterotic chiral string is indeed the heterotic ambitwistor string.

\newsec Green-Schwarz Formulation of Type II Chiral Strings

\seclab\secthree

\noindent
In this section we define the Type II Green-Schwarz chiral string from the singular gauge-fixing of the Type II Green-Schwarz action. Light-cone gauge quantization is then used to show the physical spectrum describes Type II supergravity.

%reproduces the familiar Green-Schwarz superstring when fixing the conformal gauge, and a first-order model whose fermionic constraints satisfy a SUSY-like algebra, when fixing the singular gauge.

\subsec Type II Chiral Strings

\subseclab\secthreeone

\noindent
The ${\cal{N}}=2$ generalization of the transformations \susyhet\ is given by
\eqnn \susytypeiialt
$$ \eqalignno{
\delta \theta^{\alpha} &= \epsilon^{\alpha}\ , \ \ \ \ \delta \hat{\theta}^{\hat{\alpha}} = \hat{\epsilon}^{\hat{\alpha}} \ , \ \ \ \ \delta X^{m} = -\half(\epsilon\gamma^{m}\theta) - \half(\hat{\epsilon}\gamma^{m}\hat{\theta})\cr
\delta P^{m} &= {{\cal{T}}\over 2}(\epsilon\gamma^{m}\partial_{1}\theta) - {{\cal{T}}\over 2}(\hat{\epsilon}\gamma^{m}\partial_{1}\hat{\theta})
\ , \ \ \ \ \delta e = 0 \ , \ \ \ \  \delta\bar{e} = 0 & \susytypeiialt
}
$$
where we have introduced the second fermionic coordinate $\hat{\theta}^{\hat{\alpha}}$ of the Type II superspace and used hatted indices to denote its chirality. If $\alpha$, $\hat{\alpha}$ possess the same (opposite) chirality, the superspace will be referred to as Type IIB (IIA). Furthermore, $P^{m}$ in \susytypeiialt\ stands for the actual momentum variable conjugate to $X^{m}$. The symmetries \susytypeiialt\ imply the existence of the following Type II supersymmetric invariants
\eqnn \typeiiinvariantsone
\eqnn \typeiiinvariantstwo
\eqnn \typeiiinvariantsthree
\eqnn \typeiiinvariantsfour
$$ \eqalignno{
\Pi^{m}_{i} &= \partial_{i}X^{m} + \half(\theta\gamma^{m}\partial_{i}\theta) + \half(\hat{\theta}\gamma^{m}\partial_{i}\hat{\theta}) &  \typeiiinvariantsone\cr
{\cal{P}}_{L}^{m} &= P^{m} - {\cal{T}}\partial_{1} X^{m}  - {\cal{T}}(\theta\gamma^{m}\partial_{1}\theta) & \typeiiinvariantstwo\cr
{\cal{P}}_{R}^{m} &= P^{m} + {\cal{T}}\partial_{1} X^{m}  + {\cal{T}}(\hat{\theta}\gamma^{m}\partial_{1}\hat{\theta})& \typeiiinvariantsthree\cr
{\cal{P}}^{m} &= \half({\cal{P}}_{L}^{m} + {\cal{P}}_{R}^{m}) & \typeiiinvariantsfour
} 
$$
The Type II Green-Schwarz action can then be rewritten in the more convenient way
\eqnn \GStypeiiactionalt
$$ \eqalignno{
S &= \int d^{2}\sigma \bigg[{\cal{P}}_{m}\Pi_{0}^{m} + {{\cal{T}}\over 2}\partial_{0}X^{m}[(\theta\gamma_{m}\partial_{1}\theta) - (\hat{\theta}\gamma_{m}\partial_{1}\hat{\theta})] - {{\cal{T}}\over 2}\partial_{1}X^{m}[(\theta\gamma_{m}\partial_{0}\theta) - (\hat{\theta}\gamma_{m}\partial_{0}\hat{\theta})]\cr
& -{{\cal{T}}\over 4}(\theta\gamma^{m}\partial_{0}\theta)(\hat{\theta}\gamma_{m}\partial_{1}\hat{\theta}) + {{\cal{T}}\over 4}(\theta\gamma^{m}\partial_{1}\theta)(\hat{\theta}\gamma_{m}\partial_{0}\hat{\theta})  + {e\over {\cal{T}}}{\cal{P}}_{L}^{2} + {\bar{e}\over {\cal{T}}}{\cal{P}}_{R}^{2}\bigg] & \GStypeiiactionalt
}
$$
Using algebraic manipulations similar to \varone-\vartwo, one can easily show that \GStypeiiactionalt\ is invariant under the $\kappa$-symmetry transformations
\eqnn \kappatypeii
$$ \eqalignno{
\delta \theta^{\alpha} &= (\gamma^{m}\kappa)^{\alpha}{\cal{P}}_{L\,m} \ , \ \ \  \delta\hat{\theta}^{\hat{\alpha}}  = (\gamma^{m}\hat{\kappa})^{\hat{\alpha}}{\cal{P}}_{R\,m} \ , \ \ \ \delta e = -{\cal{T}}\kappa_{\alpha}\partial_{R}\theta^{\alpha}\ , \ \ \
\delta \bar{e} = -{\cal{T}}\hat{\kappa}_{\hat{\alpha}}\partial_{L}\hat{\theta}^{\hat{\alpha}}\cr
\delta X^{m} &= \half(\delta\theta\gamma^{m}\theta) + \half(\delta\hat{\theta}\gamma^{m}\hat{\theta})\ , \ \ \ \delta P^{m} = -{{\cal{T}}\over 2}\partial_{1}(\delta\theta\gamma^{m}\theta) + {{\cal{T}}\over 2}\partial_{1}(\delta\hat{\theta}\gamma^{m}\hat{\theta}) & \kappatypeii
}
$$
where $\partial_{L} = \partial_{0} + 4\bar{e}\partial_{1}$. The action \GStypeiiactionalt\ clearly reduces to the ordinary Type II superstring action after fixing $e = \bar{e} = -{1\over 4}$, and integrating out the momentum variable $P^{m}$. If one, instead, fixes the singular gauge $e = -\bar{e} = -{1\over 4}$, one ends up with the action
%$$ \eqalignno{
%S &= \int d^{2}z \bigg[{\cal{P}}_{m}\bar{\Pi}^{m} + {{\cal{T}}\over 4}\Pi^{m}(\theta\gamma_{m}\bar{\partial}\theta) - {{\cal{T}}\over 4}\bar{\Pi}^{m}(\theta\gamma_{m}\partial\theta) - {{\cal{T}}\over 4}\Pi^{m}(\hat{\theta}\gamma_{m}\bar{\partial}\hat{\theta}) + {{\cal{T}}\over 4}\bar{\Pi}^{m}(\hat{\theta}\gamma_{m}\partial\hat{\theta})\cr
%& + {{\cal{T}}\over 8}(\theta\gamma^{m}\partial\theta)(\hat{\theta}\gamma_{m}\bar{\partial}\hat{\theta}) - {{\cal{T}}\over 8}(\theta\gamma^{m}\bar{\partial}\theta)(\hat{\theta}\gamma_{m}\partial\hat{\theta}) \bigg] &
%} 
%$$
\eqnn \typeiiactionsimp
$$ \eqalignno{
S &= \int d^{2}\sigma \bigg[P_{m}\bar{\Pi}^{m} + {{\cal{T}}\over 4}(\Pi^{m} - \bar{\Pi}^{m})(\theta\gamma_{m}\bar{\partial}\theta) - {{\cal{T}}\over 4}(\Pi^{m} - \bar{\Pi}^{m})(\hat{\theta}\gamma_{m}\bar{\partial}\hat{\theta})\cr
& + {{\cal{T}}\over 8}(\theta\gamma^{m}\partial\theta)(\hat{\theta}\gamma_{m}\bar{\partial}\hat{\theta}) - {{\cal{T}}\over 8}(\theta\gamma^{m}\bar{\partial}\theta)(\hat{\theta}\gamma_{m}\partial\hat{\theta}) \bigg] & \typeiiactionsimp
}
$$
This system is constrained by the Virasoro-like constraints: $T = {\cal{P}}_{L}^{m}{\cal{P}}_{L\,m}$, $\bar{T} = {\cal{P}}^{m}_{R}{\cal{P}}_{R\,m}$, and the fermionic constraints
\eqnn \dstypeiialt
$$ \eqalignno{
d_{\alpha} &= p_{\alpha} - \half(\gamma^{m}\theta)_{\alpha}{\cal{P}}_{L\,m} - {{\cal{T}}\over 4}(\gamma^{m}\theta)_{\alpha}(\theta\gamma_{m}\partial_{1}\theta)\cr
\hat{d}_{\hat{\alpha}} &= \hat{p}_{\hat{\alpha}} - \half(\gamma^{m}\hat{\theta})_{\hat{\alpha}}{\cal{P}}_{R\,m} - {{\cal{T}}\over 4}(\gamma^{m}\hat{\theta})_{\hat{\alpha}}(\hat{\theta}\gamma_{m}\partial_{1}\hat{\theta}) & \dstypeiialt
}
$$
whose non-trivial commutation relations are given by
\eqnn \algdstypeii
$$ \eqalignno{
\{d_{\alpha}(\sigma), d_{\beta}(\sigma')\} &= -(\gamma^{m})_{\alpha\beta}{\cal{P}}_{L\,m}\delta(\sigma-\sigma')\cr
\{\hat{d}_{\hat{\alpha}}(\sigma), \hat{d}_{\hat{\beta}}(\sigma')\} &= -(\gamma^{m})_{\hat{\alpha}\hat{\beta}}{\cal{P}}_{R\,m}\delta(\sigma-\sigma') & \algdstypeii
}
$$
and thus, they describe 16 first- and 16 second-class constraints which cannot be separated out in a Lorentz covariant manner. As is well known, this is the source of difficulties which prevents us to covariantly quantize \typeiiactionsimp. This issue will be addressed in section 5 when we introduce pure spinor variables, and a chiral free action along with a simple BRST operator are constructed. For completeness, we close this section by studying the physical spectrum of \typeiiactionsimp\ through standard light-cone gauge techniques.

\subsec Light-Cone Gauge Quantization

\subseclab\secthreetwo

As seen in eqn. \virasoroliketransf, the Type II Virasoro-like constraints will also generate gauge transformations which can be used to fix $P^{+} = k^{+}$, $X^{+} = 0$. Furthermore, the use of the $\kappa$-symmetry transformations \kappatypeii\ allows us to set $(\gamma^{+}\theta)_{\alpha} = (\gamma^{+}\hat{\theta})_{\hat{\alpha}} = 0$. The light-cone gauge fixed Type II action then reads
\eqnn \lightconetypeiiaction
$$ \eqalignno{
S &= \int d^{2}\sigma\bigg[P^{i}\bar{\partial}X^{i} + S^{a}\bar{\partial}S^{a} + \hat{S}^{\hat{a}}\bar{\partial}\hat{S}^{\hat{a}}\bigg] & \lightconetypeiiaction
}
$$
where $\hat{a}$, $\dot{\hat{a}}$ represent $SO(8)$ chiral and antichiral indices, respectively. The equations of motion for the light-cone gauge worldsheet fields imply that
\eqnn \modeexpansiontypeii
$$ \eqalignno{
X^{i} &= \sum_{n=-\infty}^{+\infty}x_{n}^{i}e^{in(\sigma_{0}-\sigma_{1})}\ \ \ \ \ , \ \ \ P^{i} = \sum_{n=-\infty}^{+\infty} p^{i}_{n}e^{in(\sigma_{0}-\sigma_{1})} \cr
 S^{a} &= \sqrt{i}\sum_{n=-\infty}^{+\infty}S^{a}_{n}e^{in(\sigma_{0} - \sigma_{1})} \ , \ \ \ \hat{S}^{\hat{a}} = \sqrt{i}\sum_{n=-\infty}^{+\infty}\hat{S}^{\hat{a}}_{n}e^{in(\sigma_{0} - \sigma_{1})} & \modeexpansiontypeii
}
$$
where the factors of $\sqrt{i}$ in \modeexpansiontypeii\ were chosen for convenience. Using the standard commutation relations
\eqnn \commutatorstwo
$$ \eqalignno{
[P^{m}(\sigma), X^{n}(\sigma')] = i\eta^{mn}\delta(\sigma - \sigma')& \ , \ \ \  \{S^{a}(\sigma), S^{b}(\sigma')\} = i\delta^{ab}\delta(\sigma-\sigma')\cr  \{\hat{S}^{\hat{a}}(\sigma), \hat{S}^{\hat{b}}(\sigma')\} &= i\delta^{\hat{a}\hat{b}}\delta(\sigma-\sigma') & \commutatorstwo
} 
$$
one can show the oscillation modes satisfy
\eqnn \pxstypeii
$$ \eqalignno{
[p_{m}^{i}, x_{n}^{j}] &= i\delta_{m,-n}\eta^{ij} \ , \ \ \ \{S^{a}_{m}, S^{b}_{n}\} = \delta_{m,-n}\delta^{ab} \ , \ \ \ \{\hat{S}^{\hat{a}}_{m}, \hat{S}^{\hat{b}}_{n}\} = \delta_{m,-n}\delta^{\hat{a}\hat{b}} & \pxstypeii
}
$$
The Virasoro-like constraints $T = {\cal{P}}_{L}^{m}{\cal{P}}_{L\,m}$, $\bar{T} = {\cal{P}}_{R}^{m}{\cal{P}}_{R\,m}$ take a simple form when written in light-cone gauge. Indeed,
\eqnn \auxseven
$$ \eqalignno{
T &= -2k^{+}[P^{-} - {\cal{T}}\partial_{1}X^{-} + {{\cal{T}}\over k^{+}}S^{a}\partial_{1}S^{a}] + (P^{i} - {\cal{T}}\partial_{1}X^{i})(P^{i} - {\cal{T}}\partial_{1}X^{i})\cr
\bar{T} &= -2k^{+}[P^{-} + {\cal{T}}\partial_{1}X^{-} - {{\cal{T}}\over k^{+}}\hat{S}^{\hat{a}}\partial_{1}\hat{S}^{\hat{a}}] + (P^{i} + {\cal{T}}\partial_{1}X^{i})(P^{i} + {\cal{T}}\partial_{1}X^{i}) & \auxseven
}
$$
Their zero modes in turn can easily be computed through eqn. \zeromodesvir\ to get
\eqnn \auxeight
$$ \eqalignno{
T_{0} &= -2p_{0}^{-}k^{+} + 2{\cal{T}}\bigg[\sum_{n=-\infty}^{+\infty}\alpha^{i}_{-n}\alpha^{i}_{n} -\sum_{n=-\infty}^{+\infty}nS^{a}_{-n}S^{a}_{n}\bigg]\cr
\bar{T}_{0} &= -2p_{0}^{-}k^{+} + 2{\cal{T}}\bigg[\sum_{n=-\infty}^{+\infty}\beta^{i}_{-n}\beta^{i}_{n} + \sum_{n=-\infty}^{+\infty}n\hat{S}^{\hat{a}}_{-n}\hat{S}^{\hat{a}}_{n}\bigg] & \auxeight
}
$$
where the oscillators $\alpha_{n}^{i}$, $\beta_{n}^{i}$ are defined as in \defalphabeta\ and satisfy the same oscillator algebras displayed in \algebraalphabeta. After making the redefinitions \redefalphabeta\ one finds that
\eqnn \Tfinaltypeii
\eqnn \barTfinaltypeii
$$ \eqalignno{
T_{0} &= -2p_{0}^{-}k^{+} + 2{\cal{T}}\alpha^{i}_{0}\alpha^{i}_{0} - 4{\cal{T}}\bigg[\sum_{n=1}^{+\infty}n:S^{a}_{-n}S^{a}_{n}: + \sum_{n=1}^{+\infty}n:\tilde{\alpha}^{i}_{-n}\tilde{\alpha}^{i}_{n}:\bigg] & \Tfinaltypeii\cr
\bar{T}_{0} &= -2p_{0}^{-}k^{+} + 2{\cal{T}}\beta^{i}_{0}\beta^{i}_{0} + 4{\cal{T}}\bigg[\sum_{n=1}^{+\infty}n:\hat{S}^{\hat{a}}_{-n}\hat{S}^{\hat{a}}_{n}: + \sum_{n=1}^{+\infty}n:\tilde{\beta}^{i}_{-n}\tilde{\beta}^{i}_{n}:\bigg] & \barTfinaltypeii
}
$$
where $:\,:$ stands for normal ordering, and the oscillators appearing in eqns. \Tfinaltypeii, \barTfinaltypeii\ obey the standard (anti)commutation relations
\eqnn \harmonicalgebrastypeii
$$ \eqalignno{
[\tilde{\alpha}^{i}_{m}, \tilde{\alpha}^{j}_{n}] &= \delta_{m,-n}\delta^{ij} \ , \ \ \ [\tilde{\beta}^{i}_{m}, \tilde{\beta}^{j}_{n}] = \delta_{m,-n}\delta^{ij}\cr
\{S^{a}_{m}, S^{b}_{n}\} &= \delta_{m,-n}\delta^{ab} \ , \ \ \ \{\hat{S}^{\hat{a}}_{m}, \hat{S}^{\hat{b}}_{n}\} = \delta_{m,-n}\delta^{\hat{a}\hat{b}} & \harmonicalgebrastypeii
}
$$
Notice that the normal ordering constants in \Tfinaltypeii, \barTfinaltypeii\ vanish since the bosonic and fermionic contributions exactly cancel each other. In this manner, one arrives at the mass formulae
\eqnn \Tfinalfinaltypeii
\eqnn \barTfinalfinaltypeii
$$ \eqalignno{
T_{0} &= -2p_{0}^{-}k^{+} + p^{i}_{0}p^{i}_{0} - 4{\cal{T}}\bigg[N_{S} + N_{\tilde{\alpha}}\bigg] & \Tfinalfinaltypeii \cr
\bar{T}_{0} &= -2p_{0}^{-}k^{+} + p^{i}_{0}p^{i}_{0} + 4{\cal{T}}\bigg[\bar{N}_{\hat{S}} + \bar{N}_{\tilde{\beta}}\bigg] & \barTfinalfinaltypeii 
}
$$
where $N_{\tilde{\alpha}}$, $\bar{N}_{\tilde{\beta}}$, $N_{S}$, $\bar{N}_{\hat{S}}$ are the number operators corresponding to the oscillator systems in eqn. \harmonicalgebrastypeii. Considering physical states are momentum eigenvectors, eqns. \Tfinalfinaltypeii, \barTfinalfinaltypeii\ imply that
\eqnn \massformulaonetypeii
\eqnn \massformulatwotypeii
$$ \eqalignno{
k^{2} - 4{\cal{T}}\bigg[N_{S} + N_{\tilde{\alpha}}\bigg] &= 0 & \massformulaonetypeii\cr
k^{2} + 4{\cal{T}}\bigg[\bar{N}_{\hat{S}} + \bar{N}_{\tilde{\beta}}\bigg] &= 0 & \massformulatwotypeii
}
$$
Consistency of eqns. \massformulaonetypeii, \massformulatwotypeii\ then requires that
\eqnn \Nconstrainttypeii
$$ \eqalignno{
N_{S} + N_{\tilde{\alpha}} + \bar{N}_{\hat{S}} +\bar{N}_{\tilde{\beta}} &= 0 & \Nconstrainttypeii
}
$$
Since the number operators possess non-negative eigenvalues, eqn. \Nconstrainttypeii\ implies that $N_{S} = N_{\tilde{\alpha}} = \bar{N}_{\hat{S}} = \bar{N}_{\tilde{\beta}} = 0$, which means the Hilbert space is only described by the massless ground state. As is well known, such a vacuum is generated by the tensor product of the states $\psi_{0} = \psi^{i} \oplus \psi^{\dot{a}}$, $\hat{\psi}_{0} = \hat{\psi}^{j} \oplus \hat{\psi}^{\dot{\hat{a}}}$ which satisfy
\eqnn \lcgeomtypeii
$$ \eqalignno{
S_{0\, a}\psi^{i} &= {1 \over \sqrt{2}}(\sigma^{i})_{a\dot{a}}\psi^{\dot{a}} \ , \ \ \ \ S_{0\, a}\psi_{\dot{a}} = {1 \over \sqrt{2}}(\sigma^{i})_{a\dot{a}}\psi_{i}\cr
\hat{S}_{0\, \hat{a}}\hat{\psi}^{i} & = {1 \over \sqrt{2}}(\sigma^{i})_{\hat{a}\dot{\hat{a}}}\hat{\psi}^{\dot{\hat{a}}}  \ , \ \ \ \ \hat{S}_{0\,\hat{a}}\hat{\psi}_{\dot{\hat{a}}} = {1 \over{\sqrt{2}}}(\sigma^{i})_{\hat{a}\dot{\hat{a}}}\hat{\psi}_{i} & \lcgeomtypeii
}
$$
and thus realize the zero mode algebras in \pxstypeii. In this manner, the quantization of the Type II action \typeiiactionsimp\ describes the Type II supergravity degrees of freedom as illustrated in Table 2, and so it matches the RNS chiral string result. Since there are no massive states in the physical spectrum, the tensionless limit does not affect the dynamics of the Hilbert space, and so tensionless Type II chiral strings describe the same physics as Type II ambitwistor strings.

\topinsert

\centerline{\vbox{%\offinterlineskip
\tabskip=0pt
\halign{ 
\vrule height2.75ex depth1.25ex width 0.6pt #\tabskip=1em &
\hfil #\hfil &\vrule # & \qquad # \hfil &\vrule # &
\hfil  #\hfil &#\vrule width 0.6pt \tabskip=0pt\cr
\noalign{\hrule height 0.6pt}
&  Mass Level && \hfil Physical States\hfil  &&  Number of States  &\cr
\noalign{\hrule}
& $M^{2} = 0$ && $(\psi^{i}\oplus \psi^{\dot{a}}) \otimes (\hat{\psi}^{j}\oplus \hat{\psi}^{\dot{\hat{a}}})$ && 128B$\,+\,$128F &\cr
\noalign{\hrule height 0.6pt}
}}}
\smallskip
{\leftskip=45pt\rightskip=20pt\baselineskip6pt\noindent\ninepoint
{\bf Table 2.} Type II SUGRA from the quantization of Type II chiral strings.
\par}
\endinsert

\newsec Pure Spinor Chiral Strings

\seclab\secfour

\noindent
In this section we construct the pure spinor version of the heterotic and Type II Green-Schwarz chiral strings by making use of the Oda-Tonin method. Integrated vertex operators are then introduced through a standard Faddeev-Popov procedure.

\subsec Heterotic Pure Spinor Chiral Strings

\subseclab\secfourone
\noindent

The Green-Schwarz constraints $d_{\alpha}$ in \dalt\ can be used to construct the BRST operator
\eqnn \psbrsthet
$$ \eqalignno{
Q &= \int dz\, \lambda^{\alpha}d_{\alpha}  & \psbrsthet
}
$$
where $\lambda^{\alpha}$ is a pure spinor variable satisfying $\lambda\gamma^{m}\lambda = 0$. Notice that its nilpotency immediately follows from the SUSY-like algebra \niceds. The BRST variation of the worldsheet fields in \GShetactionalt\ then reads
\eqnn \brsthetfields
$$ \eqalignno{
\delta \theta^{\alpha} &= \lambda^{\alpha} \ , \ \ \ \delta X^{m} = \half (\lambda\gamma^m\theta) \ , \ \ \ \delta P^m = -\half \partial_{1}(\lambda\gamma^{m}\theta) \ , \ \ \ \delta d_{\alpha} = -(\gamma^{m}\lambda)_{\alpha}{\cal{P}}_{L\,m} & \brsthetfields
}
$$
Using \brsthetfields, one easily computes the BRST variation of the action \GShetactionalt\ to be
\eqnn \brstvariationhetaction
$$ \eqalignno{
\delta S_{GS} &= \int d^2 z\, (\lambda\gamma^{m}\partial_{R}\theta){\cal{P}}_{L\,m}
& \brstvariationhetaction
}
$$
In order to make \GShetactionalt\ BRST invariant, one needs to add the extra term
\eqnn \auxten
$$ \eqalignno{
S' &= \half\int d^2 z\, (Y\gamma^{m}d)(\lambda\gamma_{m}\partial_{R}\theta) & \auxten
}
$$
where $Y_{\alpha}$ is a fixed pure spinor satisfying $\lambda Y = 1$ and it defines the pure spinor projector: $P_{\alpha}^{\beta} = \delta_{\alpha}^{\beta} - \half (\lambda\gamma^{m})_{\alpha}(Y\gamma^{m})^{\beta}$. One can now introduce the gauge-fixing fermion $\Psi = w_{\alpha}\partial_{R}\theta^{\alpha}$, where $w_{\alpha}$ is the momentum variable conjugate to $\lambda^{\alpha}$, which is defined up to the gauge transformation $\delta w_{\alpha} = (\gamma^{m}\lambda)_{\alpha}f_{m}$, for any $f_{m}$, so that the final action takes the form
\eqnn \sumofactions
$$ \eqalignno{
S &= S_{GS} + S' + Q\Psi
 & \sumofactions}
$$
where $S_{GS}$ is the standard Green-Schwarz action \GShetactionalt. Using the non-free OPE for the pure spinor variables, one readily finds
\eqnn \auxeleven
$$ \eqalignno{
Q\Psi &= d_\alpha
\partial_{R} \theta^\alpha - \half (Y\gamma^{m}d)(\lambda\gamma^{m}\partial_{R}\theta) + w_{\alpha}\partial_{R}\lambda^{\alpha} & \auxeleven
}
$$
where we used that $Q e = 0$. In this manner, the BRST-invariant heterotic action is found to be
\eqnn \pschiralhetaction
$$ \eqalignno{
S &= \int d^{2}\sigma\, \bigg[{\cal{P}}_{m}\Pi^{m}_{0} + {{\cal{T}} \over 2}\partial_{0}X^{m}(\theta\gamma_{m}\partial_{1}\theta)  - {{\cal{T}} \over 2}\partial_{1}X^{m}(\theta\gamma_{m}\partial_{0}\theta) + d_{\alpha}\partial_{0}\theta^{\alpha} + w_{\alpha}\partial_{0}\lambda^{\alpha}\cr
&+  {4e \over {\cal{T}}}\bigg({1 \over 4}{\cal{P}}^{2}_{L} - {\cal{T}}d_{\alpha}\partial_{1}\theta^{\alpha} - {\cal{T}}w_{\alpha}\partial_{1}\lambda^{\alpha}\bigg) + {4 \bar{e} \over {\cal{T}}}\bigg({1 \over 4}{\cal{P}}^{2}_{R} + T_{J}\bigg)\bigg] + S_{J} & \pschiralhetaction
}
$$
It is not hard to see that \pschiralhetaction\ reproduces the familiar pure spinor superstring action when setting the conformal gauge. If one, instead, fixes the singular gauge, one arrives at
\eqnn \pschiralhetactiontwo
$$ \eqalignno{
S &= \int d^{2}z\, \bigg[P_{m}\bar{\partial}X^{m} + p_{\alpha}\bar{\partial}\theta^{\alpha} + w_{\alpha}\bar{\partial}\lambda^{\alpha}\bigg] + S_{J} & \pschiralhetactiontwo
}
$$
which is nothing but the pure spinor ambitwistor string action proposed for the first time in \InfiniteTension. Notice that one can automatically replace $\partial_{1} \rightarrow \partial$ in $Q$ and any supersymmetric invariant due to the chiral nature of the worldsheet fields in \pschiralhetactiontwo. As in ordinary superstrings, the BRST operator \psbrsthet\ is associated to only one of the Virasoro-like constraints of the BRST-invariant action \pschiralhetaction, namely $T_{-} = {1 \over 4}{\cal{P}}_{L}^{m}{\cal{P}}_{L\,m} - {\cal{T}}d_{\alpha}\partial_{1}\theta^{\alpha} - {\cal{T}}w_{\alpha}\partial_{1}\lambda^{\alpha}$. Indeed, it was shown in \Renannnotes\ that after introducing non-minimal pure spinor variables \NMPS\ one can construct a composite operator $b_{-}$ satisfying $\{Q, b_{-}\} = T_{-}$. In this manner, the constraint $T_{+} = {1 \over 4}{\cal{P}}_{R}^{2} + T_{J}$ should enter the full BRST operator through a standard gauge-fixing procedure. Explicitly,
\eqnn \fullhetbrstop
$$ \eqalignno{
Q &= \int dz\,\bigg[\lambda^{\alpha}
d_{\alpha} + c_{+}T_{+} + b_{+}c_{+}\partial c_{+}\bigg] & \fullhetbrstop
}
$$
where $(b_{+}, c_{+})$ is the ghost system associated to the constraint $T_{+}$. This BRST operator has been considered previously in the context of sectorization \Renannsectorized. There, the pure spinor BRST charge \psbrsthet\ was obtained as a sum of two different BRST generators leaving the pure spinor action \pschiralhetactiontwo\ invariant. One of the generators was proposed in the original pure spinor ambitwistor construction \InfiniteTension, and the other one was discovered as a byproduct in an effort to extract all the superspace constraints of maximal supergravity from the Type II pure spinor ambitwistor string in a curved background \refs{\Chandiatypeii, \CHandiaonshell}. Here, we see that all of these peculiarities of the sectorized pure spinor formalism naturally arise from its relation to the Green-Schwarz formulation described in section \sectwotwo.

\medskip
The BRST cohomology can easily be inferred from the results of section \sectwothree\ and a simple replacement rule exchanging the Virasoro-like constraints by the conventional Virasoro constraints in ordinary pure spinor superstrings. In this way, the BRST-closed ghost number one vertex operator for the massless sector should take the form
\eqnn \vomasslesshet
$$ \eqalignno{
U_{SYM} &= c_{+}\lambda^{\alpha}A_{\alpha\,K}(x,\theta)J^K \ , \ \ \ \ U_{SUGRA} = c_{+}\lambda^{\alpha}A_{\alpha\,m}(x,\theta){\cal{P}}^{m}_{R} & \vomasslesshet 
}
$$
while the massive sector should mimic the vertex operator of the open superstring first massive level \nathanfirstmassive, namely
\eqnn \vomassivehet
$$ \eqalignno{
U_{massive} &= c_{+}\bigg[\partial\lambda^{\alpha}A_{\alpha}(x,\theta)  + \partial \theta^{\beta}\lambda^{\alpha} B_{\alpha\beta}(x,\theta) + :d_{\beta}\lambda^{\alpha}C^{\beta}_{\alpha}(,x\theta): \cr
& + :{\cal{P}}_{L}^{m}\lambda^{\alpha}H_{m\,\alpha}(x,\theta): + :J\lambda^{\alpha}E_{\alpha}(x,\theta): + :N^{mn}\lambda^{\alpha}F_{\alpha\,mn}(x,\theta):\bigg]& \vomassivehet
}
$$
where $N^{mn} = \half (w\gamma^{mn}\lambda)$, $J = w_{\alpha}\lambda^{\alpha}$, and $:\,:$ means normal ordering, whose definition reads
\eqnn \auxtwelve
$$ \eqalignno{
:U^{A}\lambda^{\alpha}\Phi_{\alpha\,A}(x,\theta): &= \oint {dz \over (y-z)} U^{A}(y)\lambda^{\alpha}(z)\Phi_{\alpha\,A}(x,\theta)(z) & \auxtwelve
}
$$
and $\Phi_{\alpha\, A}(x, \theta)$ are the superfields appearing in \vomassivehet. Indeed, after considering the gauge transformations generated by the image of the BRST operator and the pure spinor constraint, the vertex operators \vomasslesshet, \vomassivehet\ have been shown in  \Renannsectorized\ to correctly reproduce the superspace equations of motion of ${\cal{N}}=1$ super-Yang-Mills, ${\cal{N}}=1$ supergravity, and the first massive level of the open superstring. One more time, this remarkable fact instinctively follows here from the relation between the Green-Schwarz action \GShetchiralalt\ and the pure spinor model \pschiralhetactiontwo, and the straightforward light-cone gauge analysis carried out in section \sectwothree.

\medskip
To compute scattering amplitudes one needs to introduce integrated vertex operators. These operators have recently been constructed in \RenannChiral\ from the action of the so-called sector-splitting operator on unintegrated vertices. The origin of this operator can naturally be understood in the chiral string framework from the following comparison with the standard description of superstrings. In the conventional pure spinor heterotic superstring, the N-point correlation function for tree-level amplitudes reads
\eqnn \conventionalcorrelator
$$ \eqalignno{
{\cal{A}}_{N} &= \langle U_1(z_1)U_2(z_2) \int d^{2} z_3 V(z_3) \ldots \int d^{2} z_{N-1} V(z_{N-1}) U(z_{N}) \rangle & \conventionalcorrelator
}
$$
where $V(z)$ is the integrated vertex operator, which is related to the unintegrated one through the relation
\eqnn \uvb
$$ \eqalignno{
\{\bar{b},\{b, U\}\} &= V + Q\Omega
& \uvb}
$$
where $b$, $\bar{b}$ are the non-minimal pure spinor b-ghost and the reparametrization symmetry ghost respectively, $Q$ is the non-minimal BRST charge and $\Omega$ is a ghost number -1 operator depending on non-minimal variables. The prescription \conventionalcorrelator\ can then be inferred from a standard Faddeev-Popov procedure applied to the Green-Schwarz action \pschiralhetaction. Indeed, one can use the gauge-fixing fermions $\Psi = b F(e)$, $\bar{\Psi} = \bar{b}\bar{F}(\bar{e})$, where
\eqnn \gffs
$$ \eqalignno{
F(e) &= e + {1\over 4} + \sum_{r=1}^{N-3} s_{r}\mu_{r} \ , \ \ \ \ \bar{F}(\bar{e}) = \bar{e} + {1\over 4} + \sum_{r=1}^{N-3}\bar{s}_{r}\bar{\mu}_{r} & \gffs 
}
$$
and $\mu_{r}$, $\bar{\mu}_{r}$ are the holomorphic and antiholomorphic components of the Beltrami differential, respectively, to effectively fix the conformal gauge $e= - {1 \over 4}$, $\bar{e} = - { 1 \over 4}$ in \pschiralhetaction. In this manner, \gffs\ provides the familiar pure spinor action plus the ghost contributions
\eqnn \auxthirteen
$$ \eqalignno{
& \int d^{2}z\bigg[\bar{b}\partial \bar{c}  - \sum_{r=1}^{N-3}\bar{q}_{r}\bar{\mu}_{r}\bar{b} - \sum_{r=1}^{N-3}q_{r}\mu_{r} b \bigg] & 
\auxthirteen
}
$$
where we used that 
\eqnn \auxfourteen
$$ \eqalignno{
& Qb = T \ , \ \  Q\bar{b} = \bar{T}\ , \ \ Qs_{r} = q_{r} \ , \ \ Q e = 0 \ , \ \ Q\bar{e} = \partial \bar{c} & \auxfourteen
}
$$
and $T$, $\bar{T}$ are the usual holomorphic and antiholomorphic stress-energy tensors. After integrating out the fermionic variables $q_{r}$, $\bar{q}_{r}$, the path integral receives the contributions
\eqnn \auxfifteen
$$ \eqalignno{
& \prod_{r=1}^{N-3}(\int_{\Sigma}\mu_{r}b) \ , \ \ \ \ \ \ \  \prod_{r=1}^{N-3}(\int_{\Sigma}\bar{\mu}_{r}\bar{b}) & \auxfifteen
}
$$
which after acting on $N-3$ unintegrated vertex operators, as displayed in \uvb, reproduce the formula \conventionalcorrelator. 

\medskip
The same arguments can be used in the singular gauge. In this case, the gauge fixing fermions take the form $\Psi = b_{-}F(e)$, $\bar{\Psi} = b_{+}\bar{F}(\bar{e})$ where
\eqnn \gffssingular \foot{Notice that the particular form of the worldsheet metric in the singular gauge implies the decoupling of the antiholomorphic component of the Beltrami differential in the path integral.}
$$ \eqalignno{
F(e) &= e - {1\over 4} - \sum_{r=1}^{N-3} s_{r}\mu_{r} \ , \ \ \ \ \bar{F}(\bar{e}) = \bar{e} + {1\over 4} - \sum_{r=1}^{N-3}\bar{s}_{r}\mu_{r} & \gffssingular
}
$$
They effectively fix $e = -\bar{e} = -{1 \over 4}$ in the BRST-invariant action \pschiralhetaction, and introduce the ghost terms
\eqnn \ghostcontr
$$ \eqalignno{
& \int d^{2}z\bigg[b_{+}\bar{\partial} c_{+} - \sum_{r=1}^{N-3}s_{r} \mu_{r}T_{-} - \sum_{r=1}^{N-3}\bar{s}_{r}\mu_{r}T_{+}   - \sum_{r=1}^{N-3}\bar{q}_{r}\mu_{r} b_{+} - \sum_{r=1}^{N-3}q_{r}\mu_{r} b_{-} \bigg] & \ghostcontr
}
$$
where we used that 
\eqnn \auxsixteen
$$ \eqalignno{
& Qb_{-} = T_{-} \ , \ \  Qb_{+} = T_{+}\ , \ \ Qs_{r} = q_{r} \ , \ \ Q e = 0 \ , \ \ Q\bar{e} = \bar{\partial} c_{+} & \auxsixteen
}
$$
Recalling that $T = {1\over {\cal{T}}}(T_{+} - T_{-})$, where $T$ is the stress-energy tensor
\eqnn \chiralstressenergyt
$$ \eqalignno{
T &= -P_{m}\partial X^{m} - p_{\alpha}\partial\theta^{\alpha} - w_{\alpha}\partial\lambda^{\alpha} + T_{J}
& \chiralstressenergyt}
$$
and defining ${\cal{H}} = T_{+} + T_{-}$, the ghost contribution \ghostcontr\ can be put in the form
\eqnn \ghostcontrtwo
$$ \eqalignno{
& \int d^{2}z\bigg[b_{+}\bar{\partial} c_{+} - \sum_{r=1}^{N-3}\alpha_{r} \mu_{r}{\cal{H}}  - \sum_{r=1}^{N-3}\bar{q}_{r}\mu_{r} b_{+} - \sum_{r=1}^{N-3}q_{r}\mu_{r} b_{-} \bigg] & \ghostcontrtwo
}
$$
where we have defined
\eqnn \auxseventeen
$$ \eqalignno{
& \bar{s}_{r} + s_{r}= 2\alpha_{r} \ , \ \ \ \bar{s}_{r} - s_{r} = {2 \over {\cal{T}}}\beta_{r} & \auxseventeen
}
$$
and ignored the surface term associated to the stress-energy tensor $T$. After integrating out the fermionic and bosonic parameters $q_{r}$, $\bar{q}_{r}$, $\alpha_{r}$, the path integral receives the contributions
\eqnn \auxeighteen
$$ \eqalignno{
& \prod_{r=1}^{N-3}(\int_{\Sigma} \mu_{r}b_{-})\ , \ \ \ \ \prod_{r=1}^{N-3}(\int_{\Sigma} \mu_{r}b_{+})\ , \ \ \ \ \prod_{r=1}^{N-3}\delta (\int_{\Sigma} \mu_{r}{\cal{H}}) & \auxeighteen
}
$$
After acting on $N-3$ unintegrated vertex operators, the $b_{-}$ insertion introduces the integrated vertex operator in the pure spinor sector, the $b_{+}$ insertion removes the ghost variable $c_{+}$ in \vomasslesshet, \vomassivehet, while the delta function insertion can be interpreted as the action of a new operator. To see this, let us rewrite the delta function as
\eqnn \deltafunctionh
$$ \eqalignno{
\delta({\cal{H}}_{-1}) &= \int d \alpha_{r}e^{-\alpha_{r}{\cal{H}}_{-1}} & \deltafunctionh
}
$$
where ${\cal{H}}_{-1} = \oint dz {\cal{H}}$. The integrand of eqn. \deltafunctionh\ is nothing but the splitting operator $\Delta$ introduced in \RenannChiral\ from the requirement of BRST-closedness of integrated vertex operators. Therefore, the Green-Schwarz action \GShetactionalt\ and its gauge-fixed version yielding the pure spinor model \pschiralhetactiontwo\ provide a simple framework where several accidental features of the pure spinor chiral string become natural.

\subsec Type II Chiral Strings

\subseclab\secfourtwo

As before, the pure spinor BRST charge for the Type II superstring is constructed by making use of the Green-Schwarz constraints $d_{\alpha}$, $\hat{d}_{\hat{\alpha}}$ in \dstypeiialt\ via
\eqnn \brstchargetypeii
$$ \eqalignno{
Q &= \int dz \bigg[\lambda^{\alpha}d_{\alpha} + \hat{\lambda}^{\hat{\alpha}}\hat{d}_{\hat{\alpha}}\bigg]
 & \brstchargetypeii}
$$
where $\lambda^{\alpha}$, $\hat{\lambda}^{\hat{\alpha}}$ are pure spinor variables satisfying $(\lambda\gamma^{m}\lambda) = (\hat{\lambda}\gamma^{m}\hat{\lambda}) = 0$. The BRST variation of the worldsheet fields in \GStypeiiactionalt\ then reads
\eqnn \brstvariationtypeii
$$ \eqalignno{
\delta \theta^{\alpha} &= \lambda^{\alpha} \ , \ \ \ \delta\hat{\theta}^{\hat{\alpha}} = \hat{\lambda}^{\hat{\alpha}} \ , \ \ \ \delta d_{\alpha} = -(\gamma^{m}\lambda)_{\alpha}{\cal{P}}_{L\,m} \ , \ \ \ \delta \hat{d}_{\hat{\alpha}} = -(\gamma^{m}\hat{\lambda})_{\hat{\alpha}}{\cal{P}}_{R\,m}
\cr
\delta X^{m} &= \half (\delta\theta\gamma^{m}\theta) + \half (\hat{\theta}\gamma^{m}\hat{\theta}) \ , \ \ \ \delta P^{m} = -{{\cal{T}} \over 2}\partial_{1}(\lambda\gamma^{m}\theta) + {{\cal{T}} \over 2}\partial_{1}(\hat{\lambda}\gamma^{m}\hat{\theta})
& \brstvariationtypeii 
} 
$$
These equations imply the action \GStypeiiactionalt\ varies under BRST transformations as
\eqnn \brstvariationtypeiiaction
$$ \eqalignno{
\delta S_{GS} &= \int d^{2}\sigma\,\bigg[(\lambda\gamma^{m}\partial_{R}\theta){\cal{P}}_{L\,m} + (\hat{\lambda}\gamma^{m}\partial_{L}\hat{\theta}){\cal{P}}_{R\,m}\bigg]
& \brstvariationtypeiiaction
}
$$
In this manner, in order to make \GStypeiiactionalt\ invariant under BRST transformations, one needs to add the extra term
\eqnn \auxnineteen
$$ \eqalignno{
S' &= \half\int d^{2}\sigma \bigg[(Y\gamma^{m}d)(\lambda\gamma_{m}\partial_{R}\theta) + (\hat{Y}\gamma^{m}\hat{d})(\hat{\lambda}\gamma_{m}\partial_{L}\hat{\theta})\bigg] & \auxnineteen
}
$$
where $Y_{\alpha}$, $\hat{Y}_{\hat{\alpha}}$ are fixed pure spinors satisfying $\lambda Y = \hat{\lambda}\hat{Y} = 1$, and they define the pure spinor projectors: $P_{\alpha}^{\beta} = \delta_{\alpha}^{\beta} - \half(\lambda\gamma^{m})_{\alpha}(Y\gamma_{m})^{\beta}$, $\hat{P}_{\hat{\alpha}}^{\hat{\beta}} = \delta_{\hat{\alpha}}^{\hat{\beta}} - \half (\hat{\lambda}\gamma^{m})_{\hat{\alpha}}(\hat{Y}\gamma_{m})^{\hat{\beta}}$. One then introduces the gauge-fixing fermion $\Psi = w_{\alpha}\partial_{R}\theta^{\alpha} + \hat{w}_{\hat{\alpha}}\partial_{L}\hat{\theta}^{\hat{\alpha}}$, where $(w_{\alpha}, \hat{w}_{\hat{\alpha}})$ are the respective momenta conjugate to $(\lambda^{\alpha}, \hat{\lambda}^{\hat{\alpha}})$, which are defined up to the gauge transformations $\delta w_{\alpha} = (\gamma^{m}\lambda)_{\alpha}f_{m}$, $\hat{w}_{\hat{\alpha}} = (\gamma^{m}\hat{\lambda})_{\hat{\alpha}}\hat{f}_{m}$ for any $f_{m}$, $\hat{f}_{m}$, to get
\eqnn \sumofactionstypeii
$$ \eqalignno{
S &= S_{GS} + S' + Q\Psi & \sumofactionstypeii
} 
$$
where $S_{GS}$ is the standard Green-Schwarz action \GStypeiiactionalt. Using the non-free OPEs for the pure spinor variables, one obtains
\eqnn \purespinoractiontypeiizero
$$ \eqalignno{
S &= \int d^{2}\sigma \bigg[{\cal{P}}_{m}\Pi_{0}^{m} + {{\cal{T}}\over 2}\partial_{0}X^{m}[(\theta\gamma_{m}\partial_{1}\theta) - (\hat{\theta}\gamma_{m}\partial_{1}\hat{\theta})] - {{\cal{T}}\over 2}\partial_{1}X^{m}[(\theta\gamma_{m}\partial_{0}\theta) - (\hat{\theta}\gamma_{m}\partial_{0}\hat{\theta})]\cr
& -{{\cal{T}}\over 4}(\theta\gamma^{m}\partial_{0}\theta)(\hat{\theta}\gamma_{m}\partial_{1}\hat{\theta}) + {{\cal{T}}\over 4}(\theta\gamma^{m}\partial_{1}\theta)(\hat{\theta}\gamma_{m}\partial_{0}\hat{\theta})  + d_{\alpha}\partial_{0}\theta^{\alpha} + w_{\alpha}\partial_{0}\lambda^{\alpha}  + \hat{d}_{\hat{\alpha}}\partial_{0}\hat{\theta}^{\hat{\alpha}}  + \hat{w}_{\hat{\alpha}}\partial_{0}\hat{\lambda}^{\hat{\alpha}}\cr 
& + {4e\over {\cal{T}}}\bigg({1\over 4}{\cal{P}}_{L}^{2} - {\cal{T}}d_{\alpha}\partial_{1}\theta^{\alpha} - {\cal{T}}w_{\alpha}\partial_{1}\lambda^{\alpha}\bigg) + {4\bar{e}\over {\cal{T}}}\bigg({1\over 4}{\cal{P}}_{R}^{2} + {\cal{T}}\hat{d}_{\hat{\alpha}}\partial_{1}\hat{\theta}^{\hat{\alpha}} + {\cal{T}}\hat{w}_{\hat{\alpha}}\partial_{1}\hat{\lambda}^{\hat{\alpha}}\bigg)\bigg]
& \purespinoractiontypeiizero
}
$$
As in the heterotic case, \purespinoractiontypeiizero\ reduces to the familiar Type II pure spinor superstring action in the conformal gauge. If one, instead, fixes the singular gauge, one finds
\eqnn \purespinoractiontypeii
$$ \eqalignno{
S &= \int d^{2}z\,\bigg[P_{m}\bar{\partial}X^{m} + p_{\alpha}\bar{\partial}\theta^{\alpha} + w_{\alpha}\bar{\partial}\lambda^{\alpha} + \hat{p}_{\hat{\alpha}}\bar{\partial}\hat{\theta}^{\hat{\alpha}} + \hat{w}_{\hat{\alpha}}\bar{\partial}\hat{\lambda}^{\hat{\alpha}}\bigg]
& \purespinoractiontypeii
}
$$
The chiral nature of the action \purespinoractiontypeii\ allows us to replace $\partial_{1} \rightarrow \partial$ in $Q$ and any supersymmetric invariant. As shown in \Renannnotes, the use of non-minimal pure spinor variables allows us to introduce the composite operators $b_{-}$, $b_{+}$ satisfying
\eqnn \qbt
$$ \eqalignno{
\{Q, b_{-}\} &= T_{-} \ , \ \ \ \  \{Q , b_{+}\} = T_{+}
& \qbt
} 
$$
where $T_{-} = {1\over 4}{\cal{P}}_{L}^{2} - {\cal{T}}(d_{\alpha}\partial\theta^{\alpha} + w_{\alpha}\partial\lambda^{\alpha})$, $T_{+} = {1\over 4}{\cal{P}}_{R}^{2} + {\cal{T}}(\hat{d}_{\hat{\alpha}}\partial\hat{\theta}^{\hat{\alpha}} + \hat{w}_{\hat{\alpha}}\partial\hat{\lambda}^{\hat{\alpha}})$. 

\medskip
The BRST cohomology can also be inferred here by using our knowledge on the physical spectrum of the Green-Schwarz formulation discussed in section \secthreetwo, and the close relation between chiral and conventional pure spinor superstrings. One then proposes the massless vertex operator
\eqnn \typeiiunintegrated
$$ \eqalignno{
U_{SUGRA} &= \lambda^{\alpha}\hat{\lambda}^{\hat{\alpha}}A_{\alpha\hat{\alpha}}(x,\theta)
& \typeiiunintegrated
}
$$
That this operator describes Type II supergravity easily follows from the same analysis developed in ordinary superstrings, so we will not repeat it here. 

\medskip 
Scattering amplitudes require the presence of unintegrated and integrated vertex operators. Following the same line of reasoning discussed in section \secfourone, the gauge fixing fermions $\Psi = e + {1\over 4} + \sum_{r=1}^{N-3}s_{r}\mu_{r}$, $\bar{\Psi} = \bar{e} + {1 \over 4} + \sum_{r=1}^{N-3}\bar{s}_{r}\mu_{r}$ will give rise to the ghost contributions
\eqnn \auxtwenty
$$  \eqalignno{
& \prod_{r=1}^{N-3}(\int_{\Sigma} \mu_{r}b_{-}) \ , \ \ \ \ \prod_{r=1}^{N-3}(\int_{\Sigma} \mu_{r}b_{+})\ , \ \ \ \ \prod_{r=1}^{N-3}(\int_{\Sigma}\mu_{r}{\cal{H}}) & \auxtwenty
}
$$
where ${\cal{H}} = T_{-} + T_{+}$. In this manner, using that $A_{\alpha\hat{\alpha}}(x,\theta) = A_{\alpha}(\theta)\bar{A}_{\hat{\alpha}}(\hat{\theta})e^{ik\cdot X}$, where $A_{\alpha}(\theta)$, $\bar{A}_{\hat{\alpha}}(\hat{\theta})$ are 10D super-Yang-Mills gauge potentials, the Type II integrated vertex operator takes the form
\eqnn \typeiiintegrated
$$ \eqalignno{
V &= \int d \alpha_{r} \,e^{-\alpha_{r}{\cal{H}}_{-1}} \{b_{-}, U\}\{b_{+}, \bar{U}\}
 & \typeiiintegrated
 }
$$
One more time, one recovers the full operator proposed in \RenannChiral\ formulated from other perspectives. 
 
\subsec Integrated Vertex Operators

\subseclab\secfourthree
One can explicitly compute the integrated vertex operators discussed in previous sections by using the so-called physical operators constructed in \cederwallborninfeld\ in the superparticle context. To this end, let us first discuss the stringy realization of such operators in the construction of the standard non-minimal pure spinor $b$-ghost. Such an object was first introduced in \NMPS, and later simplified to the form (ignoring normal-ordering terms) \dynamical
\eqnn \simplifiedbghost
$$ \eqalignno{
b &= \Pi^{m}\bar{\Gamma}_{m} + {(\lambda\gamma^{mn}r)\over 4(\lambda\bar{\lambda})}\bar{\Gamma}_{m}\bar{\Gamma}_{n} - w_{\alpha}\partial\theta^{\alpha} + {(\bar{\lambda}\gamma^{m}w) \over 2(\lambda\bar{\lambda})}(\lambda\gamma^{m}\partial\theta) - s^{\alpha}\partial\bar{\lambda}_{\alpha} & \simplifiedbghost
}
$$
where $(\bar{\lambda}_{\alpha}, r_{\alpha})$ are non-minimal pure spinor variables satisfying $\bar{\lambda}\gamma^{m}\bar{\lambda} = \bar{\lambda}\gamma^{m}r = 0$, and $(\bar{w}^{\alpha}, s^{\alpha})$ are their respective conjugate momenta. In addition, the fermionic vector $\bar{\Gamma}^{m}$ in \simplifiedbghost\ is defined as
\eqnn \gammabar
$$ \eqalignno{
\bar{\Gamma}^{m} &= {(\bar{\lambda}\gamma^{m}d) \over 2(\lambda\bar{\lambda})} + {(\bar{\lambda}\gamma^{mnp}r) \over 8(\lambda\bar{\lambda})^{2}}N_{np} & \gammabar
}
$$
The object \simplifiedbghost\ can be rewritten in the more convenient way
\eqnn \bghostphysical
$$ \eqalignno{
b &= \partial\theta^{\alpha}\hat{\bf{A}}_{\alpha} - \half \bigg[\Pi^{m}\hat{\bf{A}}_{m} + d_{\alpha}\hat{\bf{W}}^{\alpha} - \half N^{mn}\hat{\bf{F}}_{mn}\bigg] - s^{\alpha}\partial\bar{\lambda}_{\alpha}&\bghostphysical
}
$$
where $(\hat{\bf{A}}_{\alpha}, \hat{\bf{A}}_{m}, \hat{\bf{W}}^{\alpha}, \hat{\bf{F}}_{mn})$ are defined by
\eqnn \physicaloperatorsone
\eqnn \physicaloperatorstwo
\eqnn \physicaloperatorsthree
\eqnn \physicaloperatorsfour
$$ \eqalignno{
\hat{\bf{A}}_{\alpha} &= {1\over 4(\lambda\bar{\lambda})}\bigg[(\gamma^{mn}\bar{\lambda})_{\alpha}N_{mn} + \bar{\lambda}_{\alpha}J\bigg] & \physicaloperatorsone \cr 
\hat{\bf{A}}_{m} &= -{(\bar{\lambda}\gamma_{m}d)\over 2(\lambda\bar{\lambda})} - {(\bar{\lambda}\gamma_{mnp}r)\over 8(\lambda\bar{\lambda})^{2}}N^{np} & \physicaloperatorstwo \cr
\hat{\bf{W}}^{\alpha} &= -{(\bar{\lambda}\gamma^{m})^{\alpha}\over 2(\lambda\bar{\lambda})}\bigg[\Pi_{m} + {(r\gamma_{m}d)\over 2(\lambda\bar{\lambda})} + {(r\gamma_{mnp}r)\over 8(\lambda\bar{\lambda})^{2}}N^{np}\bigg] & \physicaloperatorsthree\cr
\hat{\bf{F}}_{mn} &= {(\bar{\lambda}\gamma_{mnp}r)\over 4(\lambda\bar{\lambda})^{2}}\bigg[\Pi^{p} + {(r\gamma^{p}d)\over 2(\lambda\bar{\lambda})} + {(r\gamma^{pqr}r) \over 8(\lambda\bar{\lambda})^{2}}N_{qr}\bigg] & \physicaloperatorsfour
}
$$
and satisfy the relations
\eqnn \physrelationsone
\eqnn \physrelationstwo
\eqnn \physrelationsthree
\eqnn \physrelationsfour
$$ \eqalignno{
\{Q, \hat{\bf{A}}_{\alpha}\} &= -d_{\alpha} - (\gamma^{m}\lambda)_{\alpha}\hat{\bf{A}}_{m}& \physrelationsone\cr
\{Q, \hat{\bf{A}}_{m}\} &= \Pi_{m} + (\lambda\gamma_{m}\hat{\bf{W}})& \physrelationstwo\cr
\{Q, \hat{\bf{W}}^{\alpha}\} &= -{(\bar{\lambda}\gamma^{m})^{\alpha}\over 2(\lambda\bar{\lambda})}(\lambda\gamma_{m}\partial\theta) -{1\over 4}(\gamma^{mn})_{\beta}{}^{\alpha}\lambda^{\beta}\hat{\bf{F}}_{mn}& \physrelationsthree\cr
\{Q, \hat{\bf{F}}_{mn}\} &= -{(\bar{\lambda}\gamma_{mnp}r)\over 4(\lambda\bar{\lambda})^{2}}(\lambda\gamma^{p}\partial\theta) -  2(\lambda\gamma_{[m}\partial_{n]}\hat{\bf{W}}) & \physrelationsfour
}
$$
where $Q$ is the non-minimal pure spinor BRST operator. Notice the appearance of stringy effects in eqns. \physrelationsthree-\physrelationsfour\ as compared to their superparticle counterparts \cederwallborninfeld. Such effects are a direct consequence of the non-zero OPE between the variables $\Pi^{m}$ and $d_{\alpha}$ in the string worldsheet, as opposed to the vanishing commutator between $P^{m}$ and $d_{\alpha}$ in the worldline model. The use of eqns. \physrelationsone-\physrelationsfour\ allows us to readily check that $\{Q, b\} = T$, where $T = -{1 \over 2}\Pi^{2} - d_{\alpha}\partial\theta^{\alpha} - w_{\alpha}\partial\lambda^{\alpha} - \bar{w}^{\alpha}\partial\bar{\lambda}_{\alpha} - s^{\alpha}\partial r_{\alpha}$.

\medskip
As discussed in \cederwallequations, these operators act on $U = \lambda^{\alpha}A_{\alpha}$ in a very natural way. Indeed, after some algebra, one can show that
\eqnn \physactone
\eqnn \physacttwo
\eqnn \physactthree
\eqnn \physactfour
$$ \eqalignno{
\hat{\bf{A}}_{\alpha}(z)U(w) &\rightarrow {1 \over (z-w)}\bigg[A_{\alpha} - (\gamma^{m}\lambda)_{\alpha}\sigma_{m}\bigg] & \physactone\cr
\hat{\bf{A}}_{m}(z)U(w) &\rightarrow {1 \over (z-w)}\bigg[A_{m} - (\lambda\gamma_{m}\rho) + Q\sigma_{m}\bigg] & \physacttwo\cr
\hat{\bf{W}}^{\alpha}(z)U(w) & \rightarrow  {1\over (z-w)}\bigg[W^{\alpha} - Q\rho^{\alpha} + \lambda^{\alpha}s + (\gamma^{mn}\lambda)^{\alpha}s_{mn}\bigg] & \physactthree\cr 
\hat{\bf{F}}_{mn}(z)U(w) & \rightarrow {1 \over (z-w)}\bigg[F_{mn} + 4Qs_{mn} + (\lambda\gamma_{[m}g_{n]}) + (\lambda\gamma_{mn}g) \bigg] & \physactfour
}
$$
where 
\eqnn \auxtwentyone
$$ \eqalignno{
\sigma_{m} &= {(\bar{\lambda}\gamma_{m}A)\over 2(\lambda\bar{\lambda})} \ , \ \ \rho^{\alpha} = {(\gamma^{n}\bar{\lambda})_{\alpha}\over 2(\lambda\bar{\lambda})}A_{n} + {(\gamma^{n}\bar{\lambda})_{\alpha}\over 4(\lambda\bar{\lambda})^{2}}(r\gamma_{n}A) \ , \ \ s ={(\bar{\lambda}\xi) \over 4(\lambda\bar{\lambda})}\ , \ \ s_{mn} = {(\bar{\lambda}\gamma_{mn}\xi)\over 8(\lambda\bar{\lambda})}\cr 
g_{n}^{\alpha} &= -{(\gamma^{p}\bar{\lambda})^{\alpha} \over (\lambda\bar{\lambda})}\bigg[F_{pn} + 4Qs_{pn}\bigg] \ , \ \ g_{\alpha} = -{(\gamma^{pq}\bar{\lambda})_{\alpha}\over 8(\lambda\bar{\lambda})}\bigg[F_{pq} + 4Qs_{pq}\bigg] + {Qs \over 2(\lambda\bar{\lambda})} & \auxtwentyone
}
$$  and the standard 10D super-Yang-Mills equations of motion
\eqnn \tendsym
$$ \eqalignno{
D_{\alpha}A_{\beta} + D_{\beta}A_{\alpha} &= -(\gamma^{m})_{\alpha\beta}A_{m} \ , \ \ \ \ \ D_{\alpha}A_{m} = \partial_{m}A_{\alpha} + (\gamma_{m}W)_{\alpha}\cr
D_{\alpha}W^{\beta} &= -{1\over 4}(\gamma^{mn})_{\alpha}{}^{\beta}F_{mn} \, \ \ \ \ \ D_{\alpha}F_{mn} = -2(\gamma_{[m}\partial_{n]}W)_{\alpha}
& \tendsym
}
$$
were used. The terms which prevent the physical operators to reproduce the actual 10D super-Yang-Mills superfields up to BRST-exact terms are called shift-symmetry terms, and they can easily be read off from the relations \physrelationsone-\physrelationsfour. In this manner, the simple pole of $b(z)U(w)$ is easily computed to be
\eqnn \boneu
$$ \eqalignno{
b_{-1} \cdot U &= \partial\theta^{\alpha}A_{\alpha} - \Pi^{m}A_{m} - d_{\alpha}W^{\alpha} + \half N^{mn}F_{mn} & \boneu
}
$$
where we ignored BRST-exact terms, and used the identity
\eqnn \identityphys
$$ \eqalignno{
\hat{\bf{A}}_{m}\partial^{m} + \hat{\bf{W}}^{\alpha}D_{\alpha} + {1\over 4}\hat{\bf{F}}_{mn}(\lambda\gamma^{mn}\partial_{\lambda}) &= \Pi^{m}{\bf{A}}_{m} + d_{\alpha}{\bf{W}}^{\alpha} - \half N^{mn}{\bf{F}}_{mn} & \identityphys
}
$$
with $({\bf{A}}_{m}, {\bf{W}}^{\alpha}, {\bf{F}}_{mn})$ being the operator versions of $(\hat{\bf{A}}_{m}, \hat{\bf{W}}^{\alpha}, \hat{\bf{F}}_{mn})$, respectively. This is nothing but the pure spinor integrated vertex operator. A similar result has recently been obtained in \chandiabghost\ from a brute-force approach.

\medskip
The application of these ideas to the chiral case is immediate. The sectorized pure spinor $b$-ghost has been constructed in \Renannnotes\ for the Type II and heterotic cases. It differs from the conventional one in the presence of different relative numerical factors in its $r$-expansion. This is so since the OPEs of the worldsheet variables in the sectorized string differ from the conventional ones in just overall numerical factors. Using the chiral version of the physical operators defined above, namely
\eqnn \physicaloperatorsonea
\eqnn \physicaloperatorstwoa
\eqnn \physicaloperatorsthreea
\eqnn \physicaloperatorsfoura
$$ \eqalignno{
\tilde{\bf{A}}_{\alpha} &= {1\over 4(\lambda\bar{\lambda})}\bigg[(\gamma^{mn}\bar{\lambda})_{\alpha}N_{mn} + \bar{\lambda}_{\alpha}J\bigg] & \physicaloperatorsonea \cr 
\tilde{\bf{A}}_{m} &= -{(\bar{\lambda}\gamma_{m}d)\over 2(\lambda\bar{\lambda})} - {(\bar{\lambda}\gamma_{mnp}r)\over 8(\lambda\bar{\lambda})^{2}}N^{np} & \physicaloperatorstwoa \cr
\tilde{\bf{W}}^{\alpha} &= -{(\bar{\lambda}\gamma^{m})^{\alpha}\over 2(\lambda\bar{\lambda})}\bigg[{\cal{P}}_{L\,m} + {(r\gamma_{m}d)\over 2(\lambda\bar{\lambda})} + {(r\gamma_{mnp}r)\over 8(\lambda\bar{\lambda})^{2}}N^{np}\bigg] & \physicaloperatorsthreea\cr
\tilde{\bf{F}}_{mn} &= {(\bar{\lambda}\gamma_{mnp}r)\over 4(\lambda\bar{\lambda})^{2}}\bigg[{\cal{P}}_{L}^{p} + {(r\gamma^{p}d)\over 2(\lambda\bar{\lambda})} + {(r\gamma^{pqr}r) \over 8(\lambda\bar{\lambda})^{2}}N_{qr}\bigg] & \physicaloperatorsfoura
}
$$
which satisfy the relations
\eqnn \physrelationsonea
\eqnn \physrelationstwoa
\eqnn \physrelationsthreea
\eqnn \physrelationsfoura
$$ \eqalignno{
\{Q, \tilde{\bf{A}}_{\alpha}\} &= -d_{\alpha} - (\gamma^{m}\lambda)_{\alpha}\tilde{\bf{A}}_{m}& \physrelationsonea\cr
\{Q, \tilde{\bf{A}}_{m}\} &= {\cal{P}}_{L\,m} + (\lambda\gamma_{m}\tilde{\bf{W}})& \physrelationstwoa\cr
\{Q, \tilde{\bf{W}}^{\alpha}\} &= {\cal{T}}{(\bar{\lambda}\gamma^{m})^{\alpha}\over (\lambda\bar{\lambda})}(\lambda\gamma_{m}\partial\theta) -{1\over 4}(\gamma^{mn})_{\beta}{}^{\alpha}\lambda^{\beta}\tilde{\bf{F}}_{mn}& \physrelationsthreea\cr
\{Q, \tilde{\bf{F}}_{mn}\} &=  {\cal{T}}{(\bar{\lambda}\gamma_{mnp}r)\over 2(\lambda\bar{\lambda})^{2}}(\lambda\gamma^{p}\partial\theta) - 2(\lambda\gamma_{[m}\partial_{n]}\tilde{\bf{W}}) & \physrelationsfoura
}
$$
the heterotic sectorized pure spinor $b$-ghost can be compactly written as
\eqnn \bghostsectorized
$$
\eqalignno{
b_{-} &= {\cal{T}}\partial\theta^{\alpha}\tilde{\bf{A}}_{\alpha} + {1\over 4}\bigg[{\cal{P}}_{L}^{m} \tilde{\bf{A}}_{m} + d_{\alpha}\tilde{\bf{W}}^{\alpha} - \half N^{mn}\tilde{\bf{F}}_{mn}\bigg] & \bghostsectorized
}
$$
As a consistency check, one can use eqns. \physrelationsonea-\physrelationsfoura\ to show that $\{Q, b_{-}\} = T_{-}$. Thus, the integrated vertex operator, defined by the simple pole of $b_{-}(z)U(w)$, can easily be computed to be
\eqnn \integratedvertexheterotic
$$ \eqalignno{
V &= b_{-1}\cdot U = {\cal{T}}\partial\theta^{\alpha}A_{\alpha} + \half \bigg[{\cal{P}}_{L}^{m}A_{m} + d_{\alpha}W^{\alpha} - \half N^{mn}F_{mn}\bigg] & \integratedvertexheterotic
}
$$
where we ignored BRST-exact terms, and used the chiral version of \identityphys. It is not hard to see that $\{Q , V\} = \half {\cal{P}}_{L}^{m}\partial_{m}U + {\cal{T}}\partial\theta^{\alpha}D_{\alpha}U + {\cal{T}}\partial\lambda^{\alpha}A_{\alpha} = (T_{-})_{-1} \cdot U $, and so
\eqnn \qvchiral
$$ \eqalignno{
\{Q, V\} &= {{\cal{T}}\over 2}(T_{-1}\cdot U) + \half({\cal{H}}_{-1} \cdot U) & \qvchiral
}
$$
Therefore, up to a total derivative term, the BRST-variation of the integrated vertex operator \integratedvertexheterotic\ vanishes on the support of the delta function $\delta({\cal{H}}_{-1})$.

\medskip
The Type II generalization of the previous analysis straightforwardly follows from the Type II versions of the chiral physical operators \physicaloperatorsonea-\physicaloperatorsfoura. In this case, one will have two sets of such operators, namely $(\tilde{\bf{A}}_{\alpha}, \tilde{\bf{A}}_{m}, \tilde{\bf{W}}^{\alpha}, \tilde{\bf{F}}_{mn})$, $(\bar{\bf{A}}_{\hat{\alpha}}, \bar{\bf{A}}_{m}, \bar{\bf{W}}^{\hat{\alpha}}, \bar{\bf{F}}_{mn})$ which are made out of the worldsheet variables $({\cal{P}}_{L\,m}, d_{\alpha}, \lambda^{\alpha}, \ldots)$, $({\cal{P}}_{R\,m}, \hat{d}_{\hat{\alpha}}, \hat{\lambda}^{\hat{\alpha}}, \ldots)$, respectively. A computation similar to that displayed in \integratedvertexheterotic, then allows us to write the integrand of the chiral Type II integrated vertex operator \typeiiintegrated\ in the form $V =V_{L}V_{R}$, where
\eqnn \typeiichiraloperatorone
\eqnn \typeiichiraloperatortwo
$$
\eqalignno{
V_{L} &= {\cal{T}}\partial\theta^{\alpha}A_{\alpha} + \half \bigg[{\cal{P}}_{L}^{m}A_{m} + d_{\alpha}W^{\alpha} - \half N^{mn}F_{mn}\bigg] & \typeiichiraloperatorone \cr
V_{R} &= -{\cal{T}}\partial\hat{\theta}^{\hat{\alpha}}\bar{A}_{\hat{\alpha}} + \half \bigg[{\cal{P}}_{R}^{m}\bar{A}_{m} + \hat{d}_{\hat{\alpha}}\bar{W}^{\hat{\alpha}} - \half \hat{N}^{mn}\bar{F}_{mn}\bigg] & \typeiichiraloperatortwo
}
$$
where $(A_{\alpha}, A_{m}, W^{\alpha}, F_{mn})$, $(\bar{A}_{\hat{\alpha}}, \bar{A}_{m}, \bar{W}^{\hat{\alpha}}, \bar{F}_{mn})$ are two copies of the familiar 10D super-Yang-Mills superfields. One more time, one can check that up to total derivative terms, $ V = V_{L}V_{R}$ is BRST-closed on the support of the delta function $\delta({\cal{H}}_{-1})$.

\medskip
Note that when ${\cal{T}} \rightarrow 0$, the delta function insertions reduce to the ambitwistor string delta functions, namely $\bar{\delta} (P^{2})$, and the vertex operators \integratedvertexheterotic, \typeiichiraloperatorone, \typeiichiraloperatortwo\ reduce to those proposed in \InfiniteTension. It has been shown in \GomezWZA\ that this tensionless setting indeed reproduces the correct 10D super-Yang-Mills and Type II supergravity amplitudes.

\newsec Discussions and Further Directions

\seclab\secfive

\noindent
In this work we have studied the Green-Schwarz formulations of the RNS chiral strings and shown their tensionless limits indeed describe the theories predicted by ambitwistor string theories. Their pure spinor analogues were later introduced through the Oda-Tonin method, and shown to match the sectorized models proposed from different perspectives. In particular, the integrated vertex operators were easily found from a standard gauge-fixing procedure, and their forms were explicity computed by using the 10D super-Yang-Mills physical operators. 

\medskip
In \GSpurespinorstwistors\ a common origin has been proposed for the conventional Green-Schwarz and pure spinor superstrings. These models are subject to twistor-like constraints and different local gauge symmetries, so that after quantization both formulations of the superstring can be obtained from different gauge choices. In \Renannantifield, the sectorized pure spinor string was obtained after a heavy and careful antifield treatment of such a fundamental model. We expect the Green-Schwarz models studied here to arise from this same theory after choosing an appropriate gauge-fixing fermion. 

\medskip
The generalization of the chiral actions \GShetchiralalt, \typeiiactionsimp\ to curved backgrounds immediately follows from $\kappa$-symmetry. Using the curved background version of the BRST-transformations \brsthetfields, \brstvariationtypeii, the Oda-Tonin method can easily be used to obtain the pure spinor chiral actions \pschiralhetactiontwo, \purespinoractiontypeii\ in curved backgrounds. These actions have been studied in \refs{\CHandiaonshell, \Thalescurved}\ and shown to successfully couple to 10D super-Yang-Mills and ${\cal{N}}=1$ supergravity, and ${\cal{N}}=2$ supergravity for the heterotic and Type II cases, respectively. It would be interesting to use these curved background actions to study the Green-Schwarz mechanism from potential $\kappa$-symmetry and BRST anomalies, as done for ordinary superstrings in \refs{\Toninone, \Tonintwo, \ChandiaTonin}. 

\medskip
Unlike the tensionless regime, the computation of scattering amplitudes with non-zero tension appears to be a challenge due to the presence of the delta functions $\delta({\cal{H}}_{-1})$ in the path integral. However, the sign-flip trick in one of the sectors of the Green functions \Siegelchiral\ can also be used here, so that pure spinor chiral string correlators directly follow from their conventional superstring analogues. It would be interesting to apply this prescription to the scattering of heterotic chiral strings, and use the ideas of \massivepaper\ to find the manifestly supersymmetric versions of superstring amplitudes involving one single massive state.

%%%%%%%%%%%%%%%%%%%%%%%%%%%%%%%%%%%
%%%%%%%%%%%%%%%%%%%%%%%%%%%%%%%%%%%
%%%%%%%%%%%%%%%%%%%%%%%%%%%%%%%%%%%

\bigskip \noindent{\bf Acknowledgements:} I would like to thank Oliver Schlotterer for reading through an early version of this work and providing valuable comments, and Carlos Mafra for his kind support in technical issues related to TeX. I am also grateful to Henrik Johansson, Osvaldo Chandia and Renann Jusinskas for insightful discussions on related topics. This research was supported by the European Research Council under ERC-STG-804286 UNISCAMP.
%OS is indebted to Song He for insightful discussions and collaboration on related topics. 

%%%%%%%%%%%%%%%%%%%%%%%%%%%%%%%%%%%
%%%%%%%%%%%%%%%%%%%%%%%%%%%%%%%%%%%
%%%%%%%%%%%%%%%%%%%%%%%%%%%%%%%%%%%
\appendix{A}{Light-Cone Gauge Analysis of Heterotic Ambitwistor String Gravity}

%\subsec The b-Ghost On A Single Superfield

\noindent
In this Appendix we briefly review the heterotic ambitwistor string framework and study its physical spectrum in light-cone gauge.

\medskip
The heterotic ambitwistor string is defined by the action \MasonSVA
\eqnn \ambihet
$$ \eqalignno{
S &= \int d^{2}z\bigg[P_{m}\bar{\partial}X^{m} + p_{\alpha}\bar{\partial}\theta^{\alpha} + b\bar{\partial}c + \tilde{b}\bar{\partial}\tilde{c} + \psi\bar{\partial}\psi + \beta\bar{\partial}\gamma\bigg] + S_{J} & \ambihet
}
$$
and the BRST operator
\eqnn \brstambihet
$$ \eqalignno{
Q &= \oint dz\bigg[cT + bc\partial c + \gamma P\cdot \psi + \half \tilde{c}P^{2} - b\gamma^{2}\bigg] & \brstambihet
}
$$
where $T$ is the full stress-energy tensor given by
\eqnn \hettensor
$$ \eqalignno{
T &= -P\cdot \partial X - \half \psi \cdot \partial\psi + \tilde{c}\partial\tilde{b} - 2\tilde{b}\partial\tilde{c} - \half \partial\beta \gamma - {3 \over 2}\beta\partial\gamma  + T_{J}& \hettensor
}
$$
where $T_{J}$ is the stress-energy tensor associated to the current algebra system in \ambihet. The ghost systems $(b,c)$, $(\beta,\gamma)$, $(\tilde{b},\tilde{c})$ are the familiar Faddeev-Popov ghost variables corresponding to the fixing of the diffeomorphism, worldsheet supersymmetry, particle-like Hamiltonian symmetries, respectively. The BRST-cohomology of \brstambihet\ has been studied in \Berkovitslize\ and shown to be described by gauge fields satisfying the relations
\eqnn \gaugeeom
$$ \eqalignno{
\lform A_{m}^{K} - \partial_{m}(\partial \cdot A^{K}) &= 0 \ , \ \ \ \ \delta A^{K}_{m} = \partial_{m}\Lambda^{K} & \gaugeeom
}
$$
where $K$ is a Lie algebra index, spin 2-states obeying the coupled system of equations
\eqnn \spintwoeom
$$ \eqalignno{
\partial^{n}G_{mn}^{(1)} &= 0 \ , \  \partial^{n}G^{(2)}_{mn} - \partial_{m}S = 0 \ , \ -\half \lform G^{(1)}_{mn} + G^{(2)}_{mn} - {1 \over 10}\eta_{mn}G^{(2)p}_{p} = 0 \ , \ \lform S = 0\cr
\lform G^{(2)}_{mn} + 2\partial_{m}\partial_{n}S &= 0 \ , \ \ \ \delta G^{(1)}_{mn} = - \partial_{(m}\Sigma_{n)}  + {1\over 4}\eta_{mn}\partial\cdot \Sigma \ , \ \ \ \delta G^{(2)}_{mn} = -\lform \partial_{(m}\Sigma_{n)} \ , \ \ \ \delta S = 0\cr 
& &\spintwoeom
}
$$
with $\lform^{2}\Sigma = \partial_{m}(\partial\cdot \Sigma) + 2\lform \Sigma_{m} = 0$, and 2-form and 3-form fields subject to the relations
\eqnn \twoformeom
$$ \eqalignno{
\lform B_{mn} &= 0 \ , \ \ \ \ \lform C_{mnp} - \partial_{[m}B_{np]} = 0 \ , \ \ \ \  \partial^{m}B_{mn} = 0 \cr 
\partial^{m}C_{mnp} &= 0 \ , \ \ \ \ \delta B_{mn} = \partial_{[m}\Omega_{n]} \ , \ \ \ \ \delta C_{mnp} = 0 & \twoformeom
}
$$
where $\lform \Omega = \partial \cdot \Omega = 0$. Eqn. \gaugeeom\ clearly implies the existence of non-abelian gauge fields in the physical spectrum. In order to solve eqns. \spintwoeom, \twoformeom\ one can go to the light-cone reference frame where $k^{m} = (k^{+}, 0, \ldots, 0)$. Eqns. \spintwoeom\ then imply that $ \lform G^{(2)}_{mn} = 0$ for $m,n \neq -$. The component $G^{(2)}_{--}$ can be gauged away by using the transverse piece of $\bar{\Sigma}^{(0)}$ in $\Sigma_{m} = \bar{\Sigma}_{m} + (n\cdot X)\tilde{\Sigma}_{m}$. This non-plane wave expansion satisfies the e.o.m $\lform^{2} = 0$ and has been previously considered in the context of conformal supergravity \BerkovitsWitten. For convenience, we will assume that $n\cdot k = 1$, and so the only relevant component of $n^{m}$ in light-cone gauge is $n^{-}$. Notice that since $\Sigma$ is constrained to satisfy $\partial_{m}(\partial \cdot \Sigma) + 2\lform \Sigma_{m} = 0$, $\tilde{\Sigma}_{m}$ is completely determined by $\tilde{\Sigma}_{m} = -{k_{m} \over 4} (\partial \cdot \bar{\Sigma} + n \cdot \tilde{\Sigma})$. In this manner, $\lform^{2}G^{(1)}_{mn} = 0$, and thus $G^{(1)}_{mn} = \bar{G}_{mn} + (n\cdot X) \tilde{G}_{mn}$, with $\delta \bar{G}_{mn} = -\partial_{(m}\bar{\Sigma}_{n)}$ and $\partial \cdot \bar{\Sigma} = 0$. The first line of \spintwoeom\ then implies
\eqnn \relationsspintwo
$$ \eqalignno{
k^{n}\bar{G}_{mn} + n^{n} \tilde{G}_{mn} &= 0 \ , \ \ \ \ k^{n}\tilde{G}_{mn} = 0 \ , \ \ \ \ \tilde{G}_{mn} - G^{(2)}_{mn} + {1\over 10}\eta_{mn}G^{(2)p}_{p} = 0 \ , \ \ \ \ \partial^{n}G^{(2)}_{mn} = 0 & \cr
& & \relationsspintwo
}
$$
After gauging away $\bar{G}_{--}$, $\bar{G}_{-i}$, it is not hard to see that the only independent components are given by $\bar{G}_{ij}$, $G^{(2)}_{-i}$, $G^{(2)}_{ij}$, which describe 80 physical degrees of freedom. Note that $G^{(1)}_{ij}$ and $G^{(2)}_{ij}$ satisfy the same e.o.m as the spin-2 fields $G_{ij}$, $\tilde{G}_{ij}$ studied in section \sectwofour.

\medskip
Similarly, one can show that eqn. \twoformeom\ requires that $2\tilde{C}_{mnp} = \partial_{[m}B_{np]}$, where $C_{mnp} = \bar{C}_{mnp} + (n\cdot X)\tilde{C}_{mnp}$. Since $B_{mn}$ is a standard two-form field, it will describe 28 physical degrees of freedom. One can now use the first equation of the second line in \twoformeom\ to learn that $k^{m}\bar{C}_{mnp} + n^{m}\tilde{C}_{mnp} = 0$, and so $\bar{C}_{-ij}$, $\bar{C}_{ijk}$ are the only independent components of the 3-form field $C_{mnp}$. In this way, the number of physical degrees of freedom coming from the antisymmetric tensors is 112, resulting in a total of 192 bosonic degrees of freedom.

\listrefs

\bye